\documentclass[usenatbib]{mnras}
\usepackage{mathptmx}
\usepackage{breakurl}
\usepackage[T1]{fontenc}
\usepackage{bold-extra}
\usepackage{graphicx}   
\usepackage{amsmath}   
\usepackage{aecompl}
\usepackage{graphicx}	
\usepackage{amssymb}	

\newcommand{\msun} {M$_{\odot}$}
\newcommand{\src} {4U~1626--67}

\title[Broadband modelling of \src]{A broadband self-consistent 
modelling of the X-ray spectrum of \src}

\author[A. D'A\`i et al.]{A.~D'A\`i$^{1}$, \thanks{E-mail: antonino.dai@ifc.inaf.it} 
G. Cusumano$^{1}$, 
M. Del Santo$^{1}$, 
V. La Parola$^{1}$, 
A. Segreto$^{1}$
\vspace{6pt}\\
$^{1}$ INAF/IASF Palermo, via Ugo La Malfa 153, I-90146 - Palermo, Italy\\
}

\date{Accepted 2017 May 09. Received 2017 May 07; in original form 2016 December 28}
\pubyear{2016}

\begin{document}
\label{firstpage}
\pagerange{\pageref{firstpage}--\pageref{lastpage}}
\maketitle

\begin{abstract}
The  accretion-powered X-ray  pulsar \src\  is one  of the  few highly
magnetized pulsars  that accretes  through Roche-lobe overflow  from a
low-mass companion. The characteristics  of its broadband spectrum are
similar to those  of X-ray pulsars hosted in a  high-mass X-ray binary
systems, with a broad resonant  cyclotron scattering feature (CRSF) at
$\sim$\,37 keV.  In  this work, we examine the  pulse-resolved and the
pulse-averaged broadband  spectrum using  data from  \emph{NuSTAR} and
\emph{Swift}.   We  use the  Becker  \&  Wolff model  of  bulk+thermal
Comptonization  to  infer key  physical  parameters  of the  accretion
column flow and a broadband model for  the disk reflected spectrum.  In the
softer X-ray  band, we need  to add  a soft black-body  component with
$kT_{\rm  bb}$  $\sim$\,0.5  keV,  whose  characteristics  indicate  a
possible origin from the neutron star surface.  Residuals suggest that
the shape  of the cyclotron  line could be more  satisfactorily fitted
using  a narrow  core and  broader wings  and, at  higher energies,  a
second harmonic could be present at $\sim$\,61 keV.

\end{abstract}


\section{Introduction} \label{Intro}

The X-ray pulsar \src\ is hosted in a very tight binary system with 43
min orbital  period \citep{middleditch81} and  spins with a  period of
$\sim$\,7.7 s \citep{rappaport77}. Since  its discovery, it appears as
a persistent  X-ray source. Its mass  function is one of  the smallest
known  ($<$\,1.3\,$\times$\,10$^{-6}$ \msun),  thus  requiring a  very
low-mass companion.   \citet{chakrabarty98} estimated the  most likely
values for the companion's star  mass, source distance and inclination
angle being  0.08 M$_{\odot}$, $\sim$\,3 kpc  and $i\lesssim\,8$\,deg,
respectively.   However,  the same  authors  found  a larger  distance
($D$\,=\,9\,$\pm$\,4 kpc)  assuming that the observed  optical flux is
produced  by  the  accretion  disk   and  that  the  X-ray  albedo  is
$\geq$\,0.9.    This   distance   range    was   also   indicated   by
\citet{takagi16} by using  the Ghosh  \& Lamb  model to  the spin-up/down
history of \src.

The X-ray spectrum can be well  described with a combination of a soft
(temperature less than  1 keV) thermal component and  a hard power-law
component with a high-energy cut-off, typically observed at $\sim$\,20
keV. Over-imposed  to this continuum  the spectrum shows a  complex of
soft  X-ray emission  lines from  a photo-ionized  plasma at  energies
close to  1 keV, a  moderately broad iron fluorescence  emission line,
and  a broader  absorption feature  at  $\sim$\,37 keV  that has  been
interpreted    as   a    cyclotron    resonant   scattering    feature
\citep{orlandini98}.   Cyclotron lines  are  key  diagnostic tools  to
directly  infer  the strength  of  the  magnetic  field in  the  close
neighborhood of  the neutron  star (NS)  surface. These  features are
produced as an effect of resonant scattering of photons in an electron
plasma, where electrons energies are quantized according to the Landau
levels.    \citet{iwakiri12}  claimed   in  phase-resolved   spectra  of
\src\ that at the phase of the pulsed minimum the cyclotron feature could be
detected in \emph{emission} rather than in absorption.

Since  the discovery  of its  pulsed emission,  the spin  evolution of
\src\ has been  regularly monitored. Episodes of  torque reversal were
observed, with  the latest one  occurred at the  begin of 2008  from a
spin-down  to a  spin-up state  \citep{jain10,beri14}.  The  broadband
spectral  change  before and  after  the  latest torque  reversal  was
studied   by    \citet{camero-arranz12}   using    two   \emph{Suzaku}
observations. They show that the complex of the softer emission lines,
mostly  dominated by  the hydrogen-like Ne  Ly$\alpha$ line,  increased dramatically
after the torque reversal,  the softer black-body emission temperature
increased  from  0.2  to  0.5   keV,  while  the  spectral  parameters
characterizing the harder component and the cyclotron feature remained
consistent.

In this work, we report on  the broadband spectral study of the source
exploiting   the  high   energy   resolution   of  the   \emph{Nuclear
  Spectroscopic Telescope Array} \citep[\emph{NuSTAR}; ][]{harrison13}
observatory,  coupled  with  the  softer and  harder  X-ray  coverage
offered by the \emph{Swift}/X-Ray Telescope \citep[XRT; ][]{burrows05}
and \emph{Swift}/Burst  Alert Telescope  \citep[BAT; ][]{barthelmy05}.
We are able to self-consistently model the 0.5--150 keV spectrum using
a physical model and we find evidence for the
presence of a second harmonic of the CRSF at $\sim$\,61 keV.

\section{OBSERVATIONS AND DATA REDUCTION}

\subsection{\emph{NuSTAR} data reduction}
\emph{NuSTAR} observed  \src\ from  2015 May  4 12:26:07  UT to  May 5
20:41:07 UT (ObsID  30101029002), for a total collecting  time of 65.2
ks.  \emph{NuSTAR} comprises two similar Focal Plane Modules (FPMA and
FPMB) which collect photons in the 3--79 keV energy band. We extracted
high-level scientific products using \textsc{nupipeline} v.0.4.5 (part
of the \textsc{heasoft} software v.  6.19), adopting default filtering
and screening criteria.  Source events were extracted using a circular
region  of  100  arcsec  radius  centred  at  the  source  coordinates
(RA\,=\,16:32:16.79,  Dec.\,=\,-67:27:39.3). A  region  with the  same
area, but  outside the  source point-spread  function wings,  where no
other  point-like  contaminating sources  were  present,  was used  to
extract background  events.  The background-subtracted averaged count  rates in
the two modules were 15.7 and  14.2 counts s$^{-1}$ for FPMA and FPMB,
respectively.

\begin{figure}
\begin{center}
\includegraphics[height=\columnwidth, angle=-90]{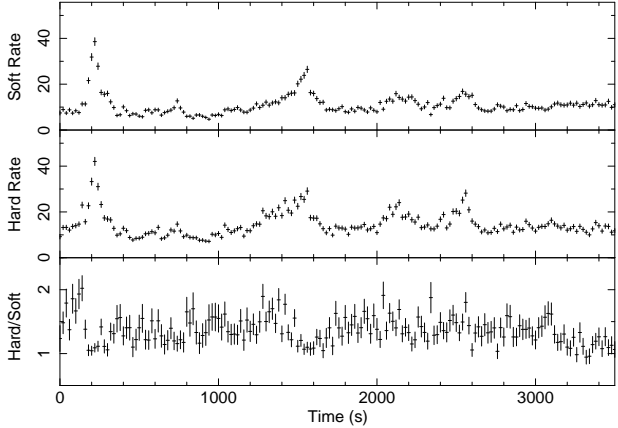}
\end{center}
\caption{From top to  bottom: soft (3.0--6.4 keV),  hard (6.4--12 keV)
  light  curves and corresponding hardness ratio for the  first 3.5  ks  of the  \emph{NuSTAR}
  observation (FPMA+FPMAB). The bin-time is 20 s.}
\label{fig:nustar}
\end{figure}

The \emph{NuSTAR}  light curve from \src\  shows moderate variability,
whose main characteristic  is the presence of an  irregular pattern of
small flares. Such flares last  tens of seconds, repeat every 100-1000
seconds, reach  a peak luminosity  which is 2--3 times  the persistent
one, and  can sometimes show complex  substructures \citep{kii86}.  To
assess the  degree of spectral  variability during these  episodes, we
selected two energy-filtered light  curves (3.0--6.4 keV and 6.4--12.0
keV,  for the  soft  and  hard bands,  respectively)  and derived  the
corresponding  hardness ratio  (HR).   The overall  level of  spectral
variability  is negligible  as it  is shown  in Fig.~\ref{fig:nustar}.
We note a marginal  softening of  the  spectrum  only during  the  peaks of  the
flaring episodes,  whereas for the  remaining part of  the observation
the  HR  variations  are not  significant  \citep[see  also][]{kii86}.
Because  the total  duration  of the  flaring peaks  is  only a  small
fraction   of  the   overall   observing  time,   we  considered   the
time-averaged spectral shape of the source only marginally affected by
these flares and proceeded on to the study the time-averaged spectrum.

\subsection{\emph{Swift} data reduction}

\emph{Swift}/XRT (XRT, hereafter) observed \src\ multiple times during
its  operational life-time.   To provide  a better  constraint of  the
softer (below  3 keV)  X-ray emission for  our broadband  analysis, we
selected  the  XRT  observation  with  identification  number  (ObsID)
00031156002  performed on  2014  March 05,  which  is the  temporarily
closest observation to  the \emph{NuSTAR} one. XRT  operated in Window
Timing mode,  and we  extracted spectra  and light  curves using  a 20
pixel  extraction  strip centred  on  the  source position  along  the
collapsed row.  Background products were  extracted from a region away
from source.  The net exposure for  this observation amounts to 4579 s
and the background-subtracted count rate is 12.5 count s$^{-1}$ in the
0.3--10 keV  range. Pile-up  is not  an issue for  WT spectra  at this
rate.

We  then used  the \emph{Swift}/BAT  (BAT, hereafter)  survey data  on
\src\ collected  between 2004 December  and 2015 November.   Data were
processed  with the  \textsc{batimager} \citep{segreto10},  a software
built  for the  analysis  of  data from  coded  mask instruments  that
performs  screening,  mosaicking  and source  detection  and  produces
scientific products of any revealed source.

\begin{figure}
\begin{center}
\includegraphics[height=\columnwidth, angle=-90]{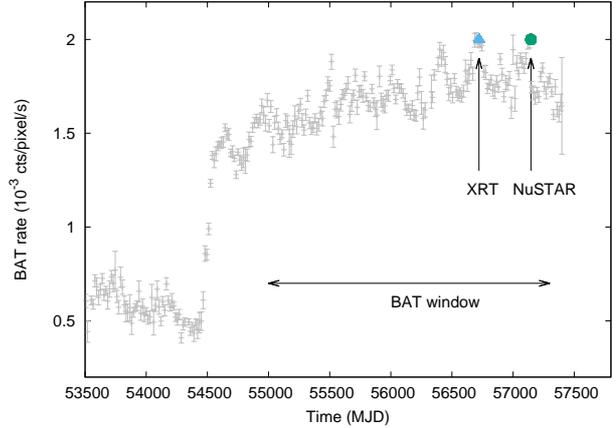}
\end{center}
\caption{\emph{Swift}/BAT light curve of  \src (15--85 keV range) with
  over-imposed  the times  of the  \emph{NuSTAR} and  \emph{Swift}/XRT
  observations considered  in this work. The  horizontal double-headed
  arrow shows the time  interval of the time-averaged \emph{Swift}/BAT
  spectrum.}
\label{fig:batlc}
\end{figure}

We show  in Fig.~\ref{fig:batlc}  the BAT  15--80 keV  long-term light
curve  of   \src\  with  over-imposed   the  dates  of  the   XRT  and
\emph{NuSTAR} observations.  The steep  increase in the  observed rate
around MJD 54500  is coincident with the time  of the torque-reversal.
After  that, the  long-term rate  appears to  be steadily  increasing,
being at  the epoch of  the \emph{NuSTAR} observation a  factor $>$\,3
higher with respect to the pre-reversal rate.

We used \textsc{xspec} v.\,12.9.0  for spectral analysis.  Spectra are
re-binned to a minimum of 20 counts per energy channel to allow use of
$\chi^2$  statistics.   Errors on  spectral  parameters  are given  at
90\,\% confidence level ($\Delta \chi^2$\,=\,2.706) unless otherwise stated.  Luminosities are
given assuming isotropic emission and a distance of 9 kpc.

\section{SPECTRAL ANALYSIS} \label{sect:spectralanalysis}

\subsection{Spectral analysis of the single datasets} \label{sect:single}

\begin{figure}
\begin{center}
\includegraphics[width=\columnwidth]{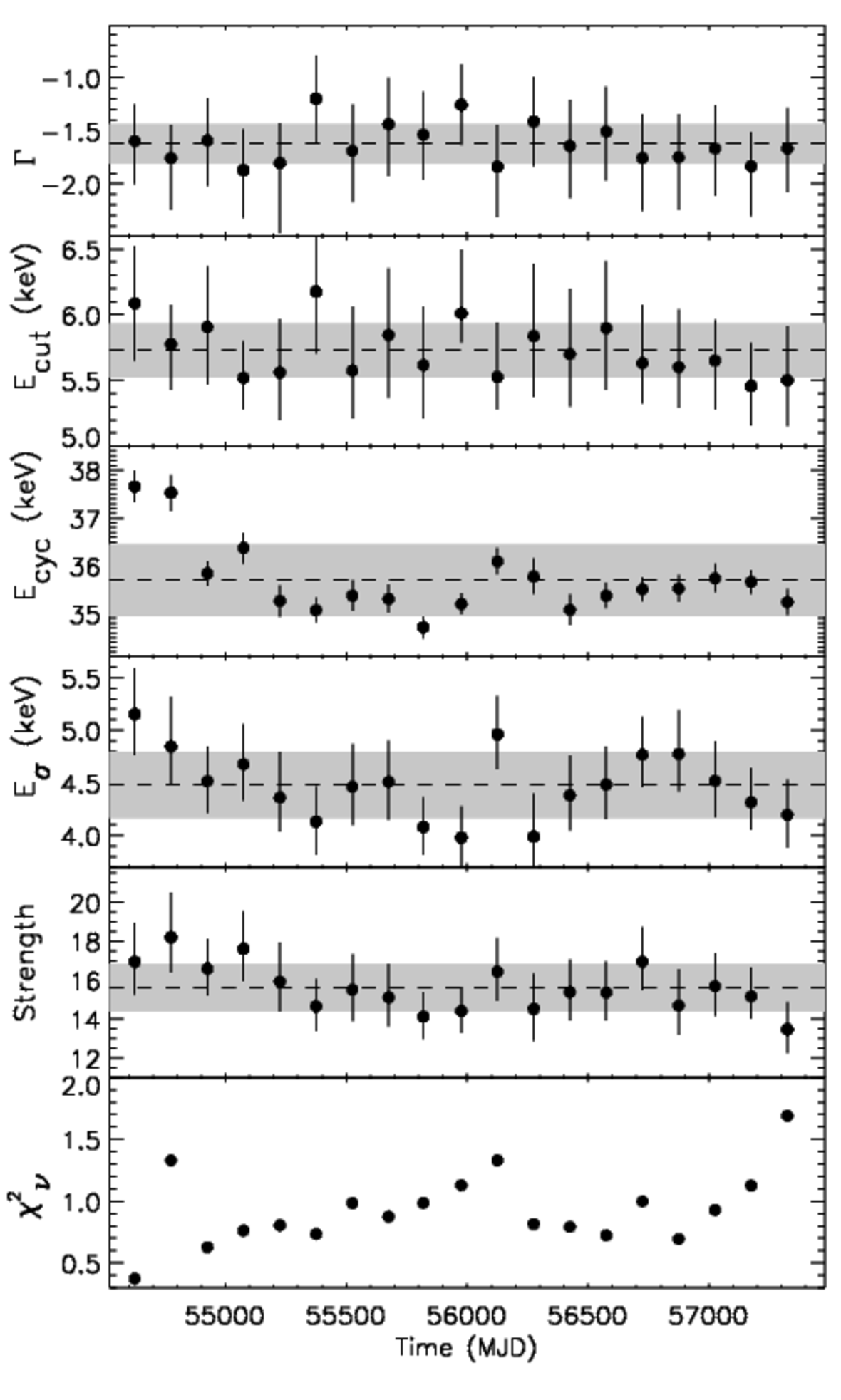}
\end{center}
\caption{Best-fit spectral parameters  of the \emph{Swift}/BAT spectra
  after the torque reversal.  Error bars are at 1\,$\sigma$ confidence
  level. Grey-shaded  areas show the standard  deviation range
    for each  parameter.  From top  to bottom: the  photon-index, the
  cut-off  energy,  the  cyclotron  line  energy,  the  width  of  the
  cyclotron line, the  strength of the line, and  the reduced $\chi^2$
  of each fit.}
\label{fig:batfits}
\end{figure}

We first focused on the spectral analysis of the \emph{Swift}/BAT data
after the  torque-reversal episode  in early  2008. Starting  from MJD
54550 (2008  March 28),  we built a  time-averaged spectrum  every 150
days, up to MJD 57400 (2016  January 13).
We assigned to each spectrum  a systematic error of 2 per cent. 
We fitted each time-averaged
spectrum  in the  15--90 keV  range  with a  cut-off power  law and  a
Gaussian  absorption feature  (\textsc{gabs} in  Xspec) at  $\sim$\,37
keV.   This simple model provided a satisfactory description of the
spectra,  with an  average  reduced $\chi^2$  of  $\sim$\,0.6, and  it
allowed  us  to check  the  spectral  variability of  the  high-energy
emission  over  this  long  period.   We  found  that  the  parameters
describing  the  continuum   emission,  that  are  in   our  case  the
photon-index  of  the  power  law  and the  cut-off  energy,  did  not
significantly  change in  the last  years.  The  line position  of the
cyclotron   line  shows   marginally  significant   (at  a   level  of
$\sim$\,3\,$\sigma$, Fig.~\ref{fig:batfits}, panel 3 from the top) 
higher  values in  the first  300 days  after the
torque  reversal,  while  after   this  period,  the  line  parameters
(position, width,  and strength) are  consistent with each  other.  We
summarize the fit results in Fig.~\ref{fig:batfits}, where we show the
best-fit  spectral values  of  each fit  as a  function  of time.   We
checked  that  the  first  two  higher line  position  values  of  the
\textsc{gabs} component are not correlated with the values of the line
width.  To this aim we repeated the fitting process, after keeping the
line  width  fixed to  the  sample  averaged  value, finding  again  a
marginal evidence  for higher  line positions value  in the  first two
time-windows. To be conservative and  to obtain a high signal-to-noise
ratio (SNR)  spectrum, we  extracted a  time-averaged \emph{Swift}/BAT
spectrum for the days 54850--57400,  where we are more confident that the
spectral shape remained  stable. In this  time window
  the source  is detected in  the 60--110 keV band  at 3.5\,$\sigma$.
Finally, we also associated to this spectrum a systematic error of the
order  of  2  per  cent   using  deviations  from  the  Crab  spectrum
\citep[see ][]{laparola16}.

We then studied the time-averaged \emph{NuSTAR}/FPMA and FPMB spectra,
leaving  a  normalization  constant  free  to  vary  between  the  two
spectra. In fitting the spectra, we used the 3--55  keV energy band
 as  the signal-to-noise  ratio (SNR) was  very low
above 55 keV.  The hard X-ray continuum emission can be well described
by different  phenomenological models usually adopted  for the spectra
of X-ray pulsars \citep[see  e.g.][]{furst13}. Because the statistical
differences  among  the  different  models were  not  significant,  to
simplify  the  presentation  of  our   results  and  without  loss  of
generality  , we  only  report on  the best-fit  model  composed of  a
power-law      modified      with      a      high-energy      cut-off
(\textsc{highecut$\times$power-law}  in  Xspec),  and a  soft  thermal
component  (\textsc{bbody}), both  absorbed  by  a neutral  absorption
column fixed at a  reference value of 1\,$\times$\,10$^{21}$ cm$^{-2}$
\citep{camero-arranz12}.  To  smooth the discontinuity at  the cut-off
energy we used a Gaussian absorption  line at the same position of the
cut-off energy as  explained in \citet{coburn02}.  We  fitted the CRSF
using  a Gaussian  absorption profile  (\textsc{gabs} in  Xspec).  The
line  is found  at 37.72\,$\pm$\,0.13  keV energy,  it has  a width  a
4.22\,$\pm$\,0.13 keV  and a strength ($S$)  of 13.6\,$\pm$\,0.7.  The
strength of the line is a measure of the optical depth ($\tau$) at the
line  centre  that  is  expressed as  $\tau=S/(\sigma  \times  \sqrt{2
  \pi})$=\,1.3\,$\pm$\,0.1. We tested if a Lorentzian profile (\textsc{cyclabs}  in Xspec) can
  also  equally well fit the cyclotron line, but we found a 
  significant worsening of the fit  ($\Delta \chi^2$\,=\,114 between the
two profiles) and, therefore, we no longer considered it.

\begin{figure}
\begin{center}
\includegraphics[height=\columnwidth, angle=-90]{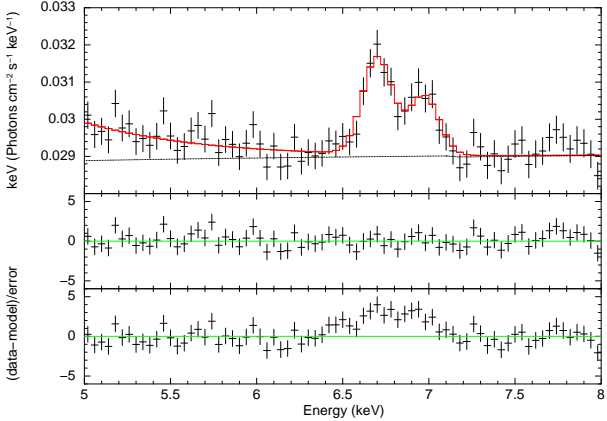}
\end{center}
\caption{Unfolded   spectrum,  model   ($Ef(E)$  representation)   and
  data/model  ratio for  the \emph{NuSTAR}  data in  the 5.0--8.0  keV
  range.   The residuals  in the  middle  (lower panel)  refer to  the
  best-fitting model that includes  (excludes) the contribution of two
  iron emission lines at fixed energies.}
\label{fig:nustariron}
\end{figure}

An apparent moderately  broad emission line in the iron  range is
significantly      detected      in       the      residuals      (see
Fig.~\ref{fig:nustariron}). Using a single emission line, we constrain
its   position  at   6.79\,$\pm$\,0.04   keV,  with   a  $\sigma$   of
160\,$\pm$\,60 eV,  and an equivalent  width of 31\,$\pm$\,6  eV.  The
line energy is not consistent  with a known rest-frame resonant energy
from  iron ions,  as it lies in-between  the energy  of the  Ly$\alpha$
transitions  of the  He-like  (6.70  keV) and  the  H-like (6.97  keV)
ion. The  line width is indicative  instead of a possible  blending of
these two lines. In fact, we can alternatively fit the residuals using
a  combination of  two lines,  fixing their  position to  the expected
rest-frame  energies (6.7  and 6.97  keV), keeping  their line  widths
tied. In this case,  we derived an upper limit to  the common width of
180 eV, thus  compatible with the emission from two  narrow lines, and
the equivalent widths are 19\,$\pm$\,6  eV and 11\,$\pm$\,5 eV for the
He-like and H-like Fe transitions, respectively.  The middle and lower
panels of  Fig.~\ref{fig:nustariron} show the residuals  for this last
fit and  for the  continuum only best-fit.   The $\chi^2$  passed from
1815 (no line, 1704 dof), to 1692 (one single, moderately broad, line,
1701 dof), to 1694 (two narrow lines at fixed energies, 1699 dof).

The   soft  thermal   black-body  component   has  a   temperature  of
0.50\,$\pm$\,0.02 keV, a  bolometric flux of 1.1\,$\times$\,10$^{-10}$
erg  cm$^{2}$  s$^{-1}$,  and  a corresponding  black-body  radius  of
11\,$\pm$\,1.5  km. The  hard X-ray  emission is  well described  by a
power-law  of  photon-index $\Gamma$\,=\,0.993\,$\pm$\,0.006,  cut-off
energy,  $E_{\rm  cut}$\,=\,21.86\,$\pm$\,0.16 keV,  e-folding  energy
$E_{\rm fold}$\,=\,10.51\,$\pm$\,0.22 keV.  The unabsorbed flux in the
extrapolated  0.1--100  keV   band  of  1.5\,$\times$\,10$^{-9}$  erg
cm$^{2}$ s$^{-1}$.  The  final reduced $\chi^2$ for this  fit is 0.996
(1701 dof).

Finally,  we examined  the  \emph{Swift}/XRT  spectrum (0.5--10.0  keV
range)  collected in  2014  March.  Adopting  a simple  two-components
continuum  consisting of  a  soft thermal  black-body  emission and  a
power-law, we  clearly detected  local residuals,  suggesting emission
structures  around,   and  below,  1   keV  and  in  the   iron  range
\citep{krauss07}.  We  used Gaussian  profiles to  model them,  and we
obtained  the  following constraints:  an  iron  line is  detected  at
6.87\,$\pm$\,0.09 keV, the line width is determined only with an upper
limit at 0.2 keV, and the  equivalent width (EW) is 120\,$\pm$\,50 eV.
This  line  is  marginally   compatible  with  a  resonant  Ly$\alpha$
transition from \ion{Fe}{xxvi}, but, in  analogy to what shown for the
\emph{NuSTAR} spectrum, a  similar fit can be obtained  by setting two
Gaussians  at  the  expected  rest-frame energies  of  the  Ly$\alpha$
transitions from \ion{Fe}{xxv} and \ion{Fe}{xxvi}.

A moderately  broadened line  is detected at  0.970\,$\pm$\,0.011 keV,
the width  is 50\,$\pm$\,17 eV, and  the EW is 70\,$\pm$\,10  eV; this
line  is  possibly  a  convolution  of  the  two  Ly$\alpha$  resonant
transitions from  \ion{Ne}{ix} (rest-frame  energy at 0.9218  keV) and
\ion{Ne}{x} (rest-frame  energy at 1.022  keV).  An equivalent  fit is
obtained by  setting two Gaussians  at these rest-frame  energies with
their widths tied.  In this fit, the EWs are 37\,$\pm$\,10 eV
and   36\,$\pm$\,8,  for   the  \ion{Ne}{ix}   and  \ion{Ne}{x}   ions
respectively.   Finally,  we  detected  the  Ly$\alpha$  \ion{O}{viii}
oxygen line at  0.64\,$\pm$\,0.02 keV, the width was kept  frozen at 0
because unconstrained,  and the  EW is 120\,$\pm$\,60  eV.  We  do not
detect emission at the energy of the \ion{O}{vii} resonant line.

\begin{figure}
\begin{center}
\includegraphics[height=\columnwidth, angle=-90]{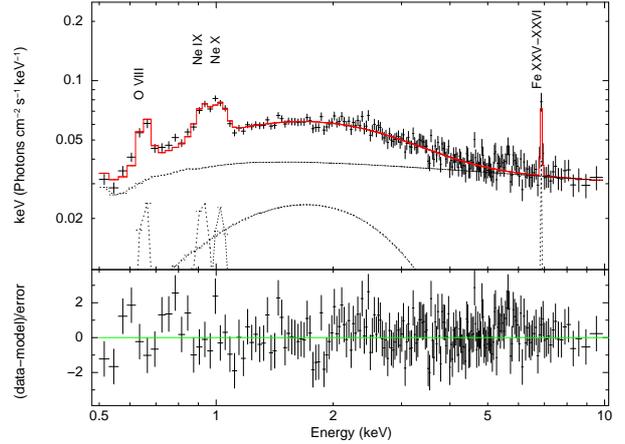}
\end{center}
\caption{Unfolded  spectrum,   model  ($Ef(E)$   representation),  and
  data/model ratio  of the \emph{Swift}/XRT spectrum  in the 0.5--10.0
  keV range. Data are visually re-binned for clarity.}
\label{fig:xrtfit}
\end{figure}

The continuum emission is well  constrained mostly at softer energies;
we     determined    an     equivalent     absorption    column     of
(6.9\,$\pm$\,3.6)\,$\times$\,10$^{20}$ cm$^{-2}$;  a thermal component
of  temperature   0.59\,$\pm$\,0.02  keV,   a  black-body   radius  of
8.0\,$\pm$\,1.1         km,          and         luminosity         of
9.0\,$\pm$\,1.0\,$\times$\,10$^{35}$   erg    s$^{-1}$;   the   harder
component  is compatible  in  this  range with  a  power-law of  index
1.16\,$\pm$\,0.10.  The  reduced $\chi^2$ for this  best-fitting model
is 0.960  (597 dof).   We show  in Fig.~\ref{fig:xrtfit}  the unfolded
spectrum, the model, the contribution  of the additive components, and
the residuals.

\subsection{The broadband spectrum of \src} \label{sect:broadband}

In this section,  we analyse the broadband,  time-averaged spectrum of
\src,  using  all the  previously  discussed  datasets, looking  for  a
self-consistent spectral model able to fit  all the data in a combined
fit.   We  leave a  normalization  constant  free  to vary  among  the
different spectra to  take into account flux differences,  as they are
different  observations performed  at  different  times, and  residual
absolute  flux  inter-calibration   uncertainties.   We  combined  and
re-binned  the  \emph{NuSTAR}  FPMA  and FPMB  spectra  into  a  single
spectrum  (FPMAB,  hereafter) to  increase  the  SNR ratio  at  higher
energies using the \textsc{addascaspec} tool, thus extending the range
up to  65.0 keV, for a  total of 168 channels.  Similarly, we
  re-binned the  original 80-channels BAT spectrum  into a 22-channels
  one. In searching for a common  model for these data, we assume to
have  only marginal  variations of  the spectral  shape of  the source
during the whole  post-torque reversal period, as shown,  at least for
the BAT  band, in Fig.~\ref{fig:batfits}.  To  describe this broadband
(0.5--150 keV)  spectrum we  adopted a physical  model suited  for the
conditions  in  the accretion  column  of  an accreting  X-ray  pulsar
\citep{becker07} and implemented as a local model in \textsc{xspec} by
\citet{ferrigno09}.  The  model (\textsc{bwmodel}) takes  into account
all the  relevant physical  processes involved in  the particle-photon
and   photon-photon    interactions   in    a   magnetically-dominated
environment. It assumes a cylindrically collimated radiation-dominated
radiative  shock above  the  NS polar  cap,  with free-fall  particles
velocities that Compton scatter photons produced by bremsstrahlung and
cyclotron processes along the  accretion column and thermal black-body
photons emerging from the accretion mound at the base of the accretion
column. Plasma  is stopped at  the shock front by  radiation pressure,
while  gas pressure  is considered  negligible. Comptonization  of the
initial  photon   spectrum  takes   into  account  both   thermal  and
bulk-motion  effects.  A  very detailed  description of  the model  is
given in \citet{wolff16}.

As            suggested             by            the            usage
guidelines\footnote{\url{http://www.isdc.unige.ch/~ferrigno/images/Documents/BW_distribution/BW_cookbook.html}},
we set the  following parameters frozen during  the fitting procedure:
the mass  and radius  of the  NS ($M_{\rm  \star}$\,=\,1.4\,\msun\ and
$R_{\rm \star}$\,=\,10 km), the source distance (9\,kpc), the magnetic
field of the NS (4.2\,$\times\,10^{12}$ G, as derived by the cyclotron
line position,  and assuming  a gravitational  redshift, $z$\,=\,0.3),
and the  normalizations of  the cyclotron and  bremsstrahlung emission
seed photon components (set to 1).  The following parameters were left
free to vary: the radius of  the accretion column, $r_0$; the electron
temperature, $T_{\textrm  e}$; the photon diffusion  parameter, $\xi$,
defined as:
\begin{equation}
\xi = \frac{\pi r_0 m_p c}{\dot{M} (\sigma_{\bot} \sigma_{\parallel} )^{1/2} },
\end{equation} 
where  $\sigma_{\bot}$  and   $\sigma_{\parallel}$  are  the  electron
cross-sections relative  to perpendicular and parallel  diffusion with
respect to the magnetic field lines,  and $m_p$ and $c$ are the proton
mass and the  speed of light; the  Comptonization parameter, $\delta$,
defined as:
\begin{equation}
\delta =  \frac{\alpha \sigma_{\parallel}}{3 \bar{\sigma}} \frac{m_{\rm e} c^2}{k_{\rm B} T_{\rm e}},
\end{equation} 
where $\bar{\sigma}$,  $m_e$, and  $k_{\rm B}$ are  the angle-averaged
electron  scattering  cross  section,   the  electron  mass,  and  the
Boltzmann constant, respectively. $\alpha$ is a factor proportional to
$\xi$ expressed as:
\begin{equation}
\alpha = 1.335 \times \frac{GM_{\star}}{R_{\star} c^2} \times \xi,
\end{equation} 

 The \textsc{bwmodel}  does not automatically  conserve energy,
  because of some  simplifying assumptions in the  energy transport in
  the post-shock  region \citep{wolff16}.  A way  to force consistency
  is to fix  the accretion rate model parameter,  $\dot{M}$, using the
  observed flux  ($F_{\rm obs}$) and  a distance guess,  inverting the
  equation $L_x=4\pi D^2 F_{\rm obs}  = G \dot{M} M_{\rm \star}/R_{\rm
    \star}$. We calculated the unabsorbed, extrapolated (0.1--150 keV) flux during the
    \emph{NuSTAR} observation $F_{\rm obs}$
  $\sim$\,1.5\,$\times$\,10$^{-9}$  erg   cm$^{-2}$  s$^{-1}$ and, for  a
  distance of  9 kpc, we  derived and fixed  in the model a $\dot{M}$
  value of  7.6\,$\times$\,10$^{16}$ g  s$^{-1}$. In  the next  section we
  will discuss how the model parameters depend on this assumption.

We  could  not  find  a  satisfactory  fit  using  only  the  absorbed
\textsc{bwmodel} for  the whole  energy band;  broad residuals  in the
softer band  suggested the need  for an additional component,  that we
modelled using  a black-body. We  then interpreted the  highly ionized
lines  described  in Sect.~\ref{sect:single}  as  due  to a  broadband
reflection  model   and,  to  self-consistently  describe   the  whole
spectrum, we adopted the  \textsc{coplrefl} reflection model developed
by \citet{ballantyne12} suited  for the typical hard  X-ray spectra of
accreting pulsars. The reflection is computed from a continuum made of
a power-law  with a  high-energy cut-off. We  froze the  parameters of
this   continuum  using   the  results   from  the   previous  section
(\emph{NuSTAR} fit)  setting the  power-law index, the  cut-off energy
and the e-folding energy to 0.995, 21.5 keV, and 12 keV, respectively.
We   allow  the   ionization  parameter   (log\,$X$)  and   the  model
normalization as free parameters.  We do not convolve the model with a
relativistic smearing  kernel, as the  inner accretion disk  radius is
expected to be truncated at a distance from the NS where no detectable
Doppler broadening  can be resolved by  our data \citep{koliopanos16}.
This model  takes well into  account the  emission from iron,  it fits
satisfactorily the oxygen  lines (small residuals are yet  seen but not
at high significance),  but leaves strong residuals  at $\sim$\,1 keV,
possibly due  to an over-abundance of  Neon with respect to  the fixed
abundance set  in the  table model  \citep[see e.g.  ][]{schulz01}; we
added   a  local   Gaussian  at   the  fixed   energy  0.97   keV  and
$\sigma$\,=\,0.05 keV to flatten residuals in this range.

We describe  the cyclotron line  with a Gaussian profile.  However, as
noted by  the different  values in  the line  parameters shown  in the
previous section, the FPMAB and  BAT spectra cannot be fitted together
with tied parameters. The most significant differences were the values
of line energy and line depth. We retain that the energy shift is most
likely due to a gain drift of  the BAT response matrix with respect to
the  FPMAB's one  \citep[see Fig.~2  in ][]{baumgartner13},  while the
difference  in the  depth values  (of the  order of  20--30 per  cent)
reflects  a  small  difference  between the  two  averaged  fluxes  as
measured in the  two different time windows.  Therefore,  we applied a
gain  correction to  the BAT  spectrum and  we left  the depth  of the
cyclotron line in  the BAT spectrum as a free  parameter.  We obtained
for the BAT  matrix an offset of $\sim$\,1.2  keV \citep[as similarly
 found by ][]{ferrigno16,  doroshenko17} with  respect to FPMAB,  while the
slope was consistent with unity.

\begin{figure*}
\begin{center}
\includegraphics[width=15cm]{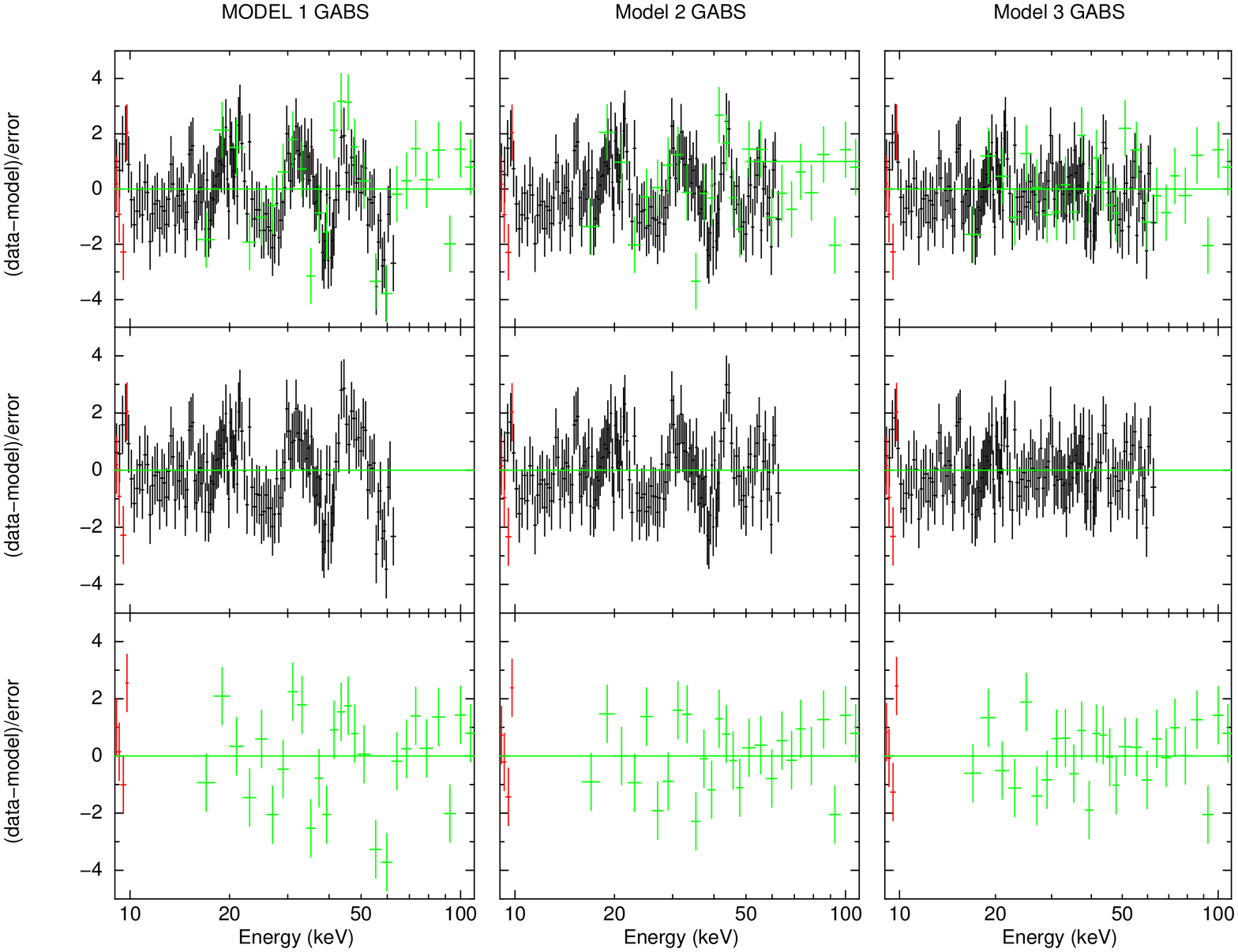}
\end{center}
\caption{Residuals  in units  of $\sigma$  for the  XRT+FPMAB+BAT data
  (upper  panels),  XRT+FPMAB  (middle  panels),  and  XRT+BAT  (lower
  panels),  using the  \textsc{bwmodel} with  one Gaussian  absorption
  line at 38 keV (first column),  with two Gaussians at $\sim$\,38 and
  61 keV (second  column) and with two nested  Gaussians at $\sim$\,38
  and one at 61  keV (third column). FPMAB data in  black, BAT data in
  green, XRT data in red. We show, for clarity's sake, only the 9--110
  keV range.}
\label{fig:fits1}
\end{figure*}

Further, we noted  that the highest energy emission, above and around
the cyclotron line,  gave still significant residuals both  in the BAT
and FPMAB data  (see upper panel of  Fig.~\ref{fig:fits1}). To resolve
the nature of these residuals,  if either due to unresolved systematic
mismatch between  the $NuSTAR$ and  \emph{Swift} spectra, or to  a bad
modelling of the continuum, or to a missing component, we obtained the
best-fits separately  for the XRT+FPMAB  and for the  XRT+BAT spectra.
As shown by the  residuals in 
Fig.~\ref{fig:fits1},   the  general   unsatisfactory  value   of  the
$\chi^2$, is  not linked to  the combined FPMAB  and BAT fits,  but
rather suggests  the need for  a more complex continuum,  or cyclotron
line  description,  or  for  an additional  component.   Secondly,  we
compared  the  residuals  pattern  with  the  one  obtained  from  the
phenomenological description using the \textsc{highecut} model for the
hard   X-ray  emission. In  this   case,  we   swapped  the
  \textsc{bwmodel}  component for  the \textsc{highecut},  leaving the
  parameters of the softer black-body and \textsc{coplrefl} components
  free to  vary accordingly.  The \textsc{highecut} model  offered a
slightly different pattern  for the residuals, but it  also pointed to
the  presence of  an absorption  feature around  60\,keV. The  overall
quality of the fit  was unsatisfactory (reduced $\chi^2$\,=\,1.34, for
788 dof) as
in the case of the \textsc{bwmodel}.  We concluded that the physically
consistent  description  of  the  continuum  offered  a  statistically
similar  value compared  with  the most  widely used  phenomenological
description, and  it is therefore  not directly accountable  by itself
for the unsatisfactory value of the $\chi^2$.

Subsequently, we tested the statistical significance for the addition of a
second absorption  Gaussian profile  (model \textsc{bwm2}),  which was
strongly suggested both by the  residuals in the \textsc{highecut} and
\textsc{bwmodel}. This second absorption Gaussian line could represent
a higher cyclotron  harmonic and to better constrain  its position and
depth, we tied  the width of the  line to that of  the first harmonic.
We found that the improvement of the $\chi^2$ due to the flattening of
the  residuals  in the  higher  energy  range is  always  significant,
obtaining  a $\Delta  \chi^2$\,=\,40, 52,  and 116,  for the  XRT+BAT,
XRT+FPMAB, and XRT+BAT+FPMAB fits, respectively (\textsc{bwmodel}).

Finally,  we  found  that  residuals could  be  flattened  around  the
cyclotron line energies  using a more complex line  profile, but still
preserving its symmetric characteristic, by imposing the superposition
of two Gaussians with a common line energy, but free widths and depths
(model  \textsc{bwm3}).  We  note, however,  that  this  more
  complex  profile   for  the  fundamental  cyclotron   line  is  only
  marginally  seen  in  the  FPMAB  spectrum  alone,  but  it  becomes
  statistically  very   significant  in   combination  with   the  XRT
  spectrum.  We cannot,  therefore, rule  out that  the shape  of this
  cyclotron feature  is biased by  the inter-calibration of  these two
  spectra, that  are not contemporaneous.  We  comparatively report in
  Table~\ref{tab:fits} the final best-fitting parameter values for the
  three  models, where  it  can  be seen  that  the  modelling of  the
  fundamental cyclotron line only marginally impacts the determination
  of the other fit parameters.  For completeness, we also report
  the  best-fit parameters  adopting the  phenomenological  model
  \textsc{highecut}  (Table~\ref{tab:fits}, last  column).

\begin{table*}
\caption{Fits results.   We show the best-fitting  parameters of three
  models that have a  common continuum (\textsc{bwmodel}, as discussed
  in Sect.~\ref{sect:broadband}), and differ  for the presence of one,
  two,   and   three   absorption  features   (models   \textsc{bwm1},
  \textsc{bwm2}, and \textsc{bwm3},  respectively).  The \textsc{bwm1}
  model models  the known  CRSF at  $\sim$\,37 keV;  the \textsc{bwm2}
  model  adds  a  possible  second harmonic  at  $\sim$\,61  keV;  the
  \textsc{bwm3} model adds another Gaussian  line at the same position
  energy  of  the first  CRSF.   In  the  last  column we  report  the
  best-fitting  parameter  using  a   phenomenological  model  with  a
  power-law  with  a  high-energy cut-off  (model  \textsc{highecut}).
  Fluxes are  calculated in  the 0.1--100  keV band  and refer  to the
  \emph{NuSTAR} spectrum, for which  the intercalibration constant was
  fixed to one. $C_{\rm bat}$ and $C_{\rm xrt}$ are the multiplicative
  model   normalization  constants   for   the  \emph{Swift}/BAT   and
  \emph{Swift}/XRT spectra.}
\begin{tabular}{llll|l}
\hline
                                                         & \textsc{bwm1} & \textsc{bwm2}  & \textsc{bwm3} & \textsc{highecut}\\
\hline
$N_{\rm H}$  (10$^{21}$ cm$^{-2}$)                         & 0.5$\pm$0.2 & 0.6$\pm$0.2                     & 0.4$\pm$0.2 & 1.5$\pm$0.15 \\

\textsc{bbody} $kT$ (keV)                                & 0.513$\pm$0.011   & 0.548$\pm$0.011                   & 0.532$\pm$0.013 & 0.569$\pm$0.014\\
\textsc{bbody} $R_{\rm bb}$ (km)                         & 3.70$\pm$0.20     & 9.6$\pm$0.5                       & 10.7$\pm$2.6    & 2.8$\pm$0.2\\
\textsc{bbody} Flux (10$^{-10}$ erg cm$^{-2}$ s$^{-1}$)  & 1.2              & 1.10                              & 1.15            & 1.02 \\

\textsc{coplrefl} log\,($X$)                              & 2.6$\pm$0.1 & 3.2$\pm$0.1 & 3.2$\pm$0.1 & 2.6$\pm$0.1 \\
\textsc{coplrefl} Flux (10$^{-10}$ erg cm$^{-2}$ s$^{-1}$)& 1.2 & 0.8 & 0.8                         & 1.1 \\

\textsc{bwmodel} $\xi$                                   & 1.18$\pm$0.02         & 1.31$\pm$0.05           & 1.21$_{-0.05}^{+0.08}$ & \\
\textsc{bwmodel} $\delta$                                & 3.12$\pm$0.18         & 2.3$_{-0.3}^{+0.2}$     & 2.9$\pm$0.4            & \\
\textsc{bwmodel} $T_{\rm e}$  (keV)                      & 4.05$\pm$0.09         & 4.7$\pm$0.20            & 4.5$\pm$0.2            & \\
\textsc{bwmodel} $R_{\rm 0}$  (m)                        & 24.8$\pm$0.8          & 28.5$^{+1.3}_{-1.0}$    & 26.0$^{+2.0}_{-1.5}$   & \\

\textsc{bwmodel} Flux (10$^{-10}$ erg cm$^{-2}$ s$^{-1}$)& 14.5                  & 15.8 & 16.4 & \\

\textsc{highecut} $\Gamma$ & & &                                   & 0.98$^{+0.01}_{-0.03}$ \\ 
\textsc{highecut} $E_{\rm fold}$ (keV) & & &                       & 12.1$\pm$0.5 \\ 
\textsc{highecut} $E_{\rm cut}$  (keV) & & &                       & 22.0$\pm$0.5\\ 
\textsc{highecut} Flux (10$^{-10}$ erg cm$^{-2}$ s$^{-1}$) & & &   & 15.7  \\ 

\textsc{gabs} $E_{\rm line}$ (keV)                                & 37.4$\pm$0.12  & 38.00$\pm$0.16 & 37.95$\pm$0.15     & 37.95$\pm$0.15 \\
\textsc{gabs} Width (keV)                                         & 4.88$\pm$0.10  & 5.34$\pm$0.16  & 6.0$\pm$0.3        & 3.9$_{-0.6}^{+0.3}$  \\
\textsc{gabs} Strength                                            & 20.9$\pm$0.7   & 25.6$\pm$1.3   & 23.0$\pm$0.9       & 11.0$_{-2.7}^{+2.2}$ \\

\textsc{gabs$_2$} $E_{\rm line}$                             &      & 60.8$_{-0.9}^{+1.1}$ & 61.0$\pm$1.0             & 67$\pm$3\\
\textsc{gabs$_2$} Strength                                   &      & 19$_{-4}^{+5}$       & 22$\pm$5                 & 50$_{-10}^{+17}$\\

\textsc{gabs$_3$} Width (keV)                                & &                           & 2.9$\pm$0.6         & 7.8$\pm$0.6\\
\textsc{gabs$_3$} Strength                                   & &                           & 3.7$_{-1.2}^{+2.1}$ & 12$\pm$3.3 \\

$C_{\rm bat}$              & 0.828$\pm$0.011       & 0.823$\pm$0.010 & 0.815$\pm$0.011  & 0.811$\pm$0.010   \\
$C_{\rm xrt}$              & 1.217$\pm$0.016       & 1.222$\pm$0.014 & 1.219$\pm$0.016  & 1.221$\pm$0.016  \\

$\chi^2_{\rm red}$ (dof)   & 1.271 (789) & 1.133 (787)     & 1.051 (785)      &  1.048 (783) \\
\hline
\end{tabular}
\label{tab:fits}
\end{table*}

\begin{figure*}
\begin{center}
\includegraphics[angle=-90, width=2\columnwidth]{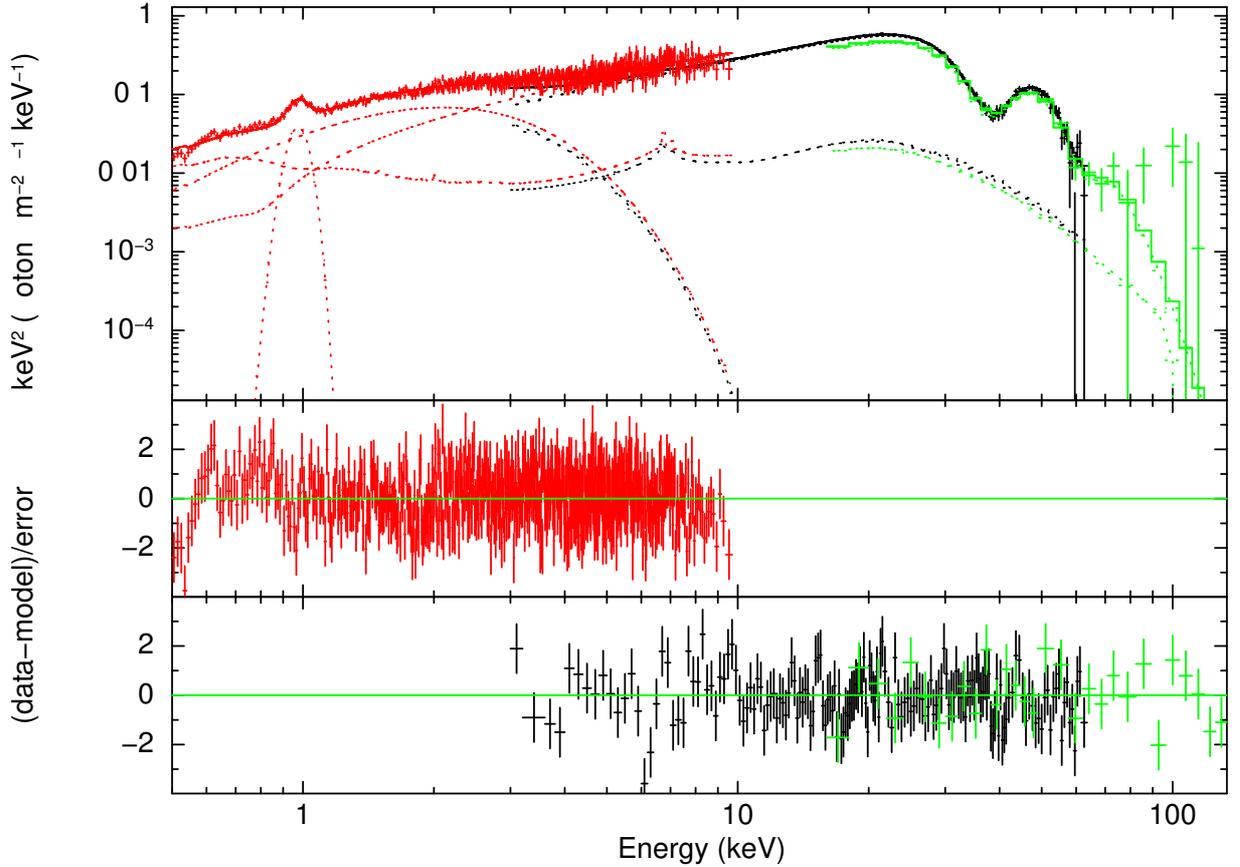}
\end{center}
\caption{Data,   unfolded   model   and  residuals   for   the 
  \textsc{bwm3} model.  XRT, FPMAB and BAT data in red, black, and 
  gree, respectively.}
\label{fig:fits3}
\end{figure*}

\subsection{Dependence of the \textsc{bwmodel} parameters on the distance estimate}

\begin{figure}
\begin{center}
\includegraphics[width=\columnwidth]{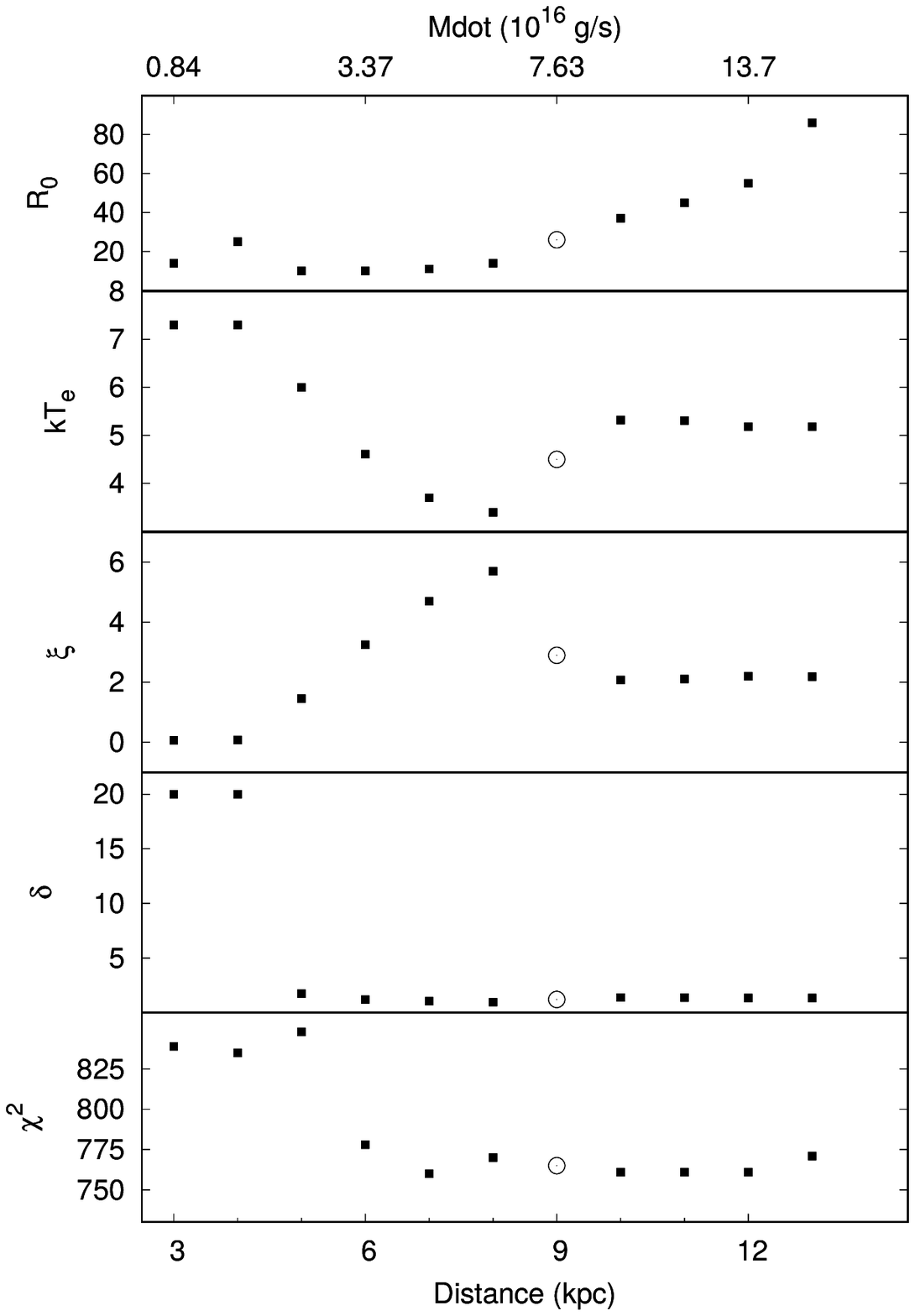}
\end{center}
\caption{Best-fit  parameter  values  of  the  \textsc{bwmodel}  as  a
  function  of the  source distance.   From top  to bottom:  the $R_0$
  column radius  (m), the  electron temperature $kT_{\rm e}$ (keV), $\xi$,  $\delta$, and
  the fit  $\chi^2$. The  open circle data point refers to a
  distance of 9 kpc as in Table~\ref{tab:fits}.}
\label{fig:distance}
\end{figure}

As discussed in the previous section, to force consistency between the
energy implied by  the \textsc{bwmodel} and that corresponding to the
observed flux ($F_{\rm obs}$ 
  $\sim$\,1.5\,$\times$\,10$^{-9}$  erg   cm$^{-2}$  s$^{-1}$),  
  we set  the $\dot{M}$ parameter  fixed. However,
this choice  depends on the assumed  distance, which, for the  case of
\src\  is  highly  uncertain.  Therefore,  we  fitted  the  data  with
\textsc{BWM3} model (the  BAT spectrum was not  considered to simplify
the  fitting procedure)  assuming  distance  values, and  consequently
different $\dot{M}$,  within the range  3--13 kpc at  a step of  1 kpc
(see  Fig.~\ref{fig:distance}).    For  distances  $D<$\,6   kpc,  the
$\chi^2$ value  is much worse than  for any larger distance.   Some of
the  \textsc{bwmodel} parameters  tend to  the extreme  values of  the
allowed parameter  space ($\delta$ to  the upper hard limit,  $\xi$ to
the   bottom  limit)   and  the   physical  interpretation   could  be
unreliable. For $6  \leq D < 9$, we obtained  acceptable fits, but the
electron  temperature  and  the  $\xi$  parameters  resulted  strongly
anti-correlated.   For distance  values $D\geq  9$, parameters  do not
show any significant correlation: $kT_e$,  $\delta$, and $\xi$ reach a
plateau  value  and  the  only  parameter  that  shows  a  significant
correlation with the distance is $R_0$.  Besides the distance
  issue,  the  accretion  rate   estimation  depends  on  two  further
  approximations we made: the whole observed flux is dissipated in the
  shock region  and the  emission comes from  only one  pole. However,
  these   represent  second-order   approximations  compared   to  the
  uncertainty on the distance.

\section{Spin-resolved spectral analysis of the \emph{NuSTAR} data}

\begin{figure}
\begin{center}
\includegraphics[height=\columnwidth, angle=-90]{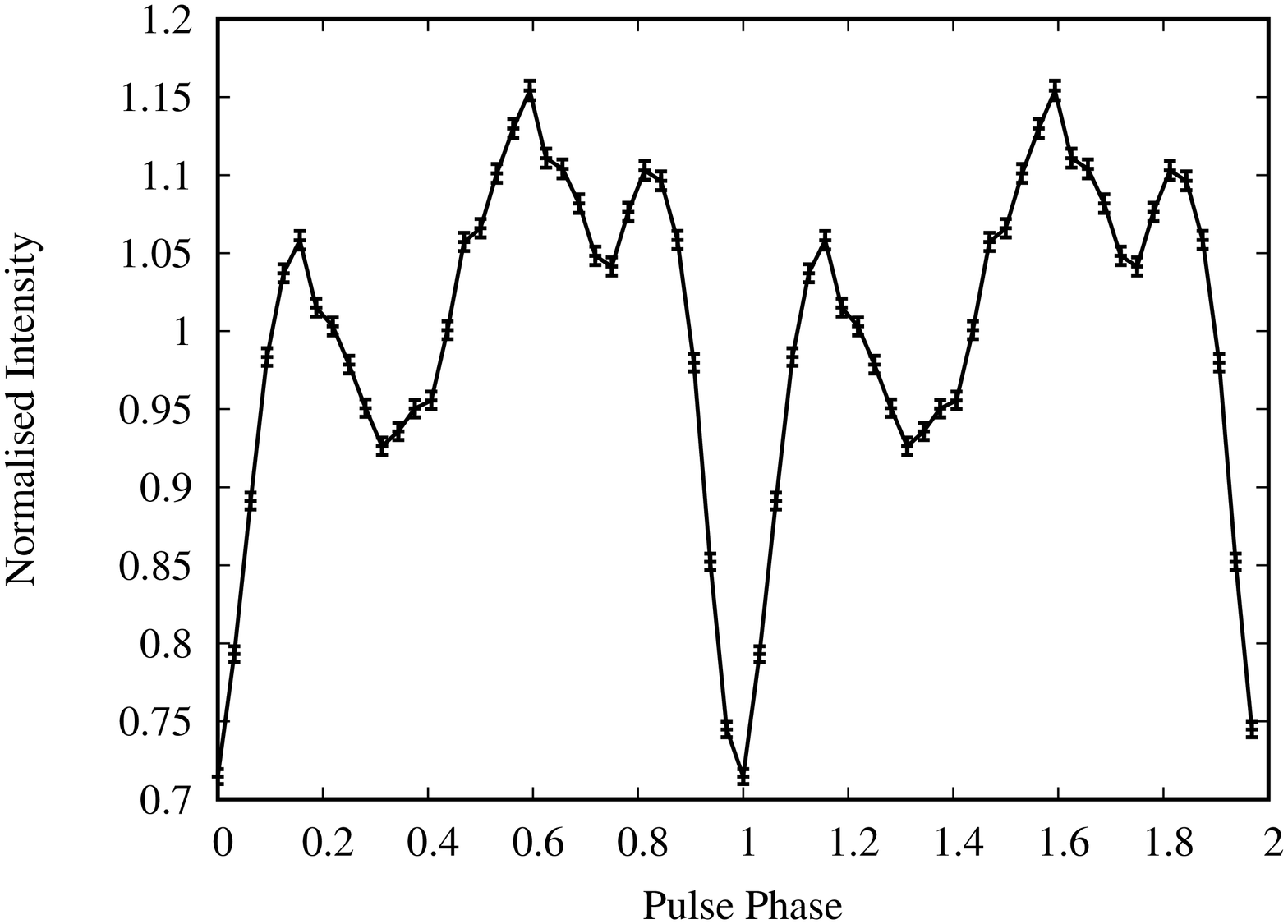}
\includegraphics[height=\columnwidth, angle=-90]{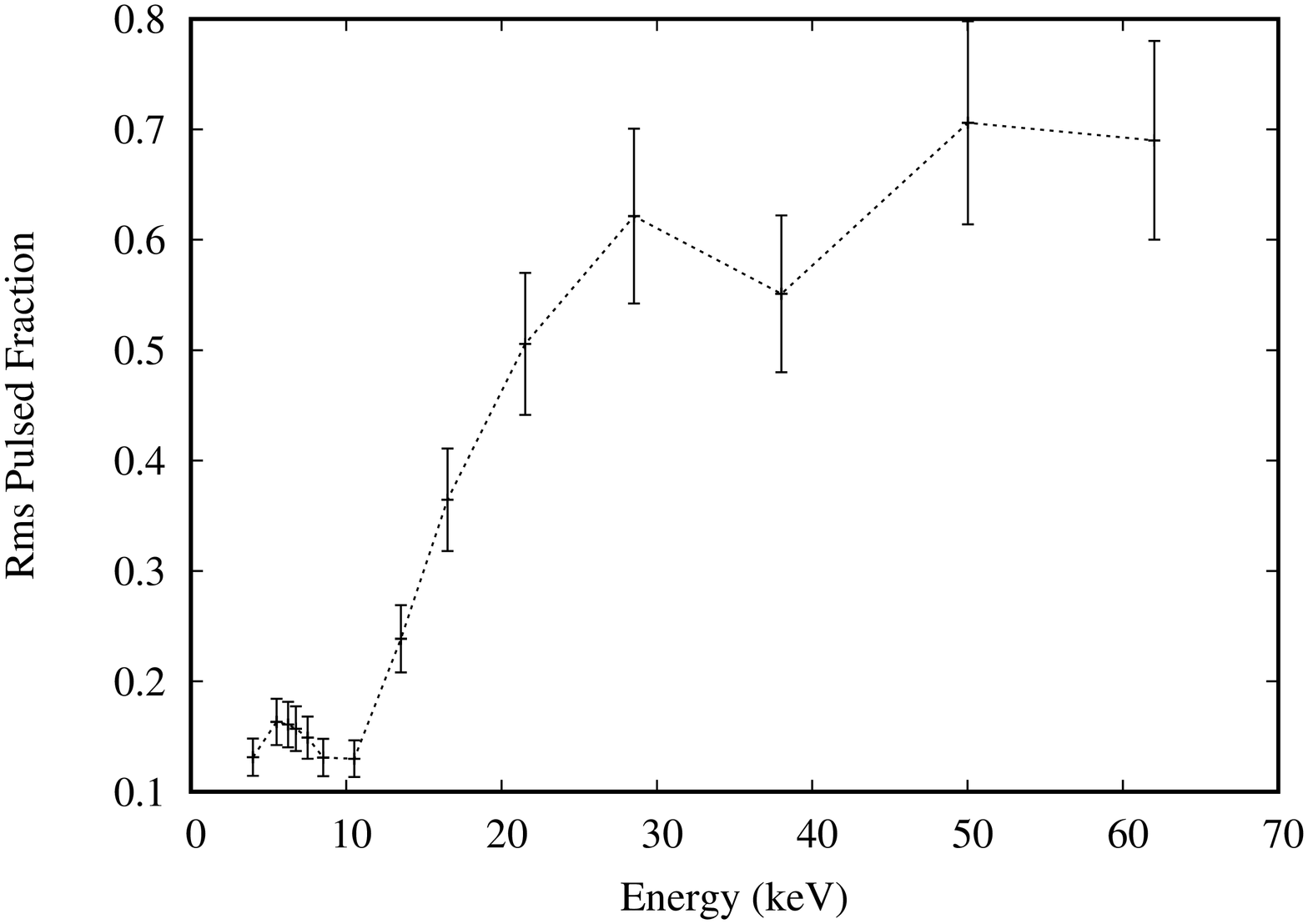}
\end{center}
\caption{Upper  panel:  folded  profile  of  the  pulsed  emission  of
  \src\  in the  whole \emph{NuSTAR}  band.  Lower panel:  rms pulsed fraction versus energy.}
\label{fig:folded}
\end{figure}

We performed  a timing  analysis of  the pulsed  emission of  \src. We
barycentred the photon arrival times  with respect to the Solar system
barycentre  using the  \textsc{barycorr} tool.   Adopting the  folding
search technique (\textsc{efsearch}  tool), we found a  spin period of
7.672952(3) s,  consistent with the \emph{Fermi}/Gamma-ray  Burst Monitor
(GBM)
value\footnote{\url{http://gammaray.nsstc.nasa.gov/gbm/science/pulsars/lightcurves/4u1626.html}}.
The background-subtracted folded pulsed profile  is shown in the upper
panel of Fig.~\ref{fig:folded},  while in the lower panel  we show the
energy-dependence of the  pulsed fraction ($P_{\rm rms}$)  as given by
the  root-mean-squared value  (rms) of  the pulsed  profile. 
  \citet{an15}  showed that  this  definition of  $P_{\rm  rms}$ is  a
  reliable  estimator of  the \textit{true}  value, although  a direct
  comparison with other  estimators needs a proper scale  factor to be
  taken into  account. We verified  that the $P_{\rm  rms}$ value of  the folded
  profile is  consistent within  a few percent  with what  obtained by
  computing  the  truncated  Fourier  expansion  and  subtracting  the
  Fourier noise \citep[Eq.\,3 in ][]{an15}.

\begin{figure}
\begin{center}
\includegraphics[width=\columnwidth]{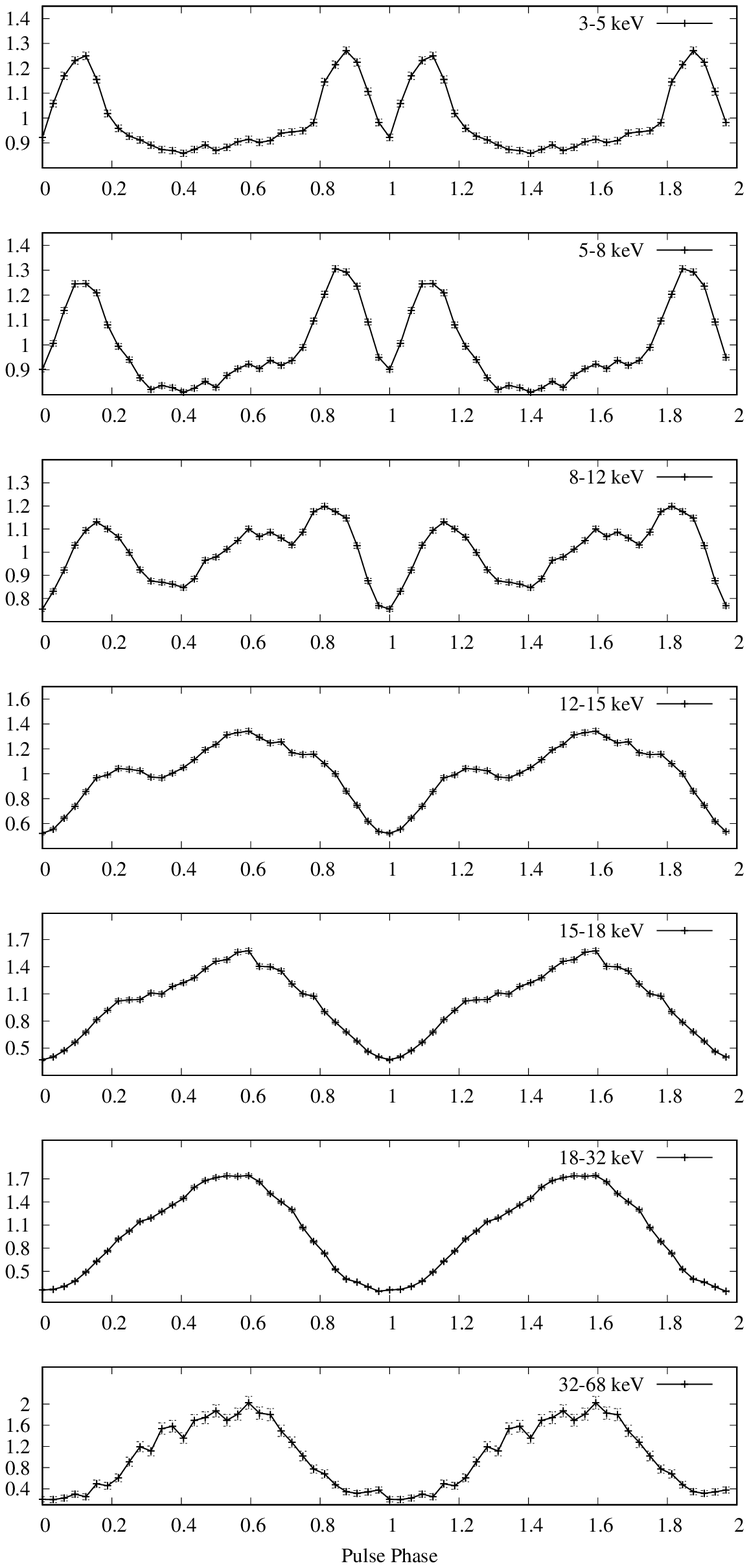}
\end{center}
\caption{Energy-resolved  folded  profiles in  selected  \emph{NuSTAR}
  energy bands. The $y$-axis shows the normalized intensity.}
\label{fig:folded2}
\end{figure}

The pulsed  fraction has a very  steep increase above 10  keV, and the
plot of  the energy-resolved profiles in  Fig.~\ref{fig:folded2} shows
that this  increase is also  marked with  a significant change  in the
pulsed  emission  from a  double-peaked  profile,  for energies  below
$\sim$\,12 keV, to  a single-peaked profile at  higher energies.  Such
results are  consistent with a previous  analysis on the shape  of the
pulsed emission  after the most recent  torque reversal \citep{beri14}
and indicate that  during this last spin-up phase the  profile has not
significantly varied.

\begin{figure*}
\begin{tabular}{cc}
\includegraphics[width=\columnwidth]{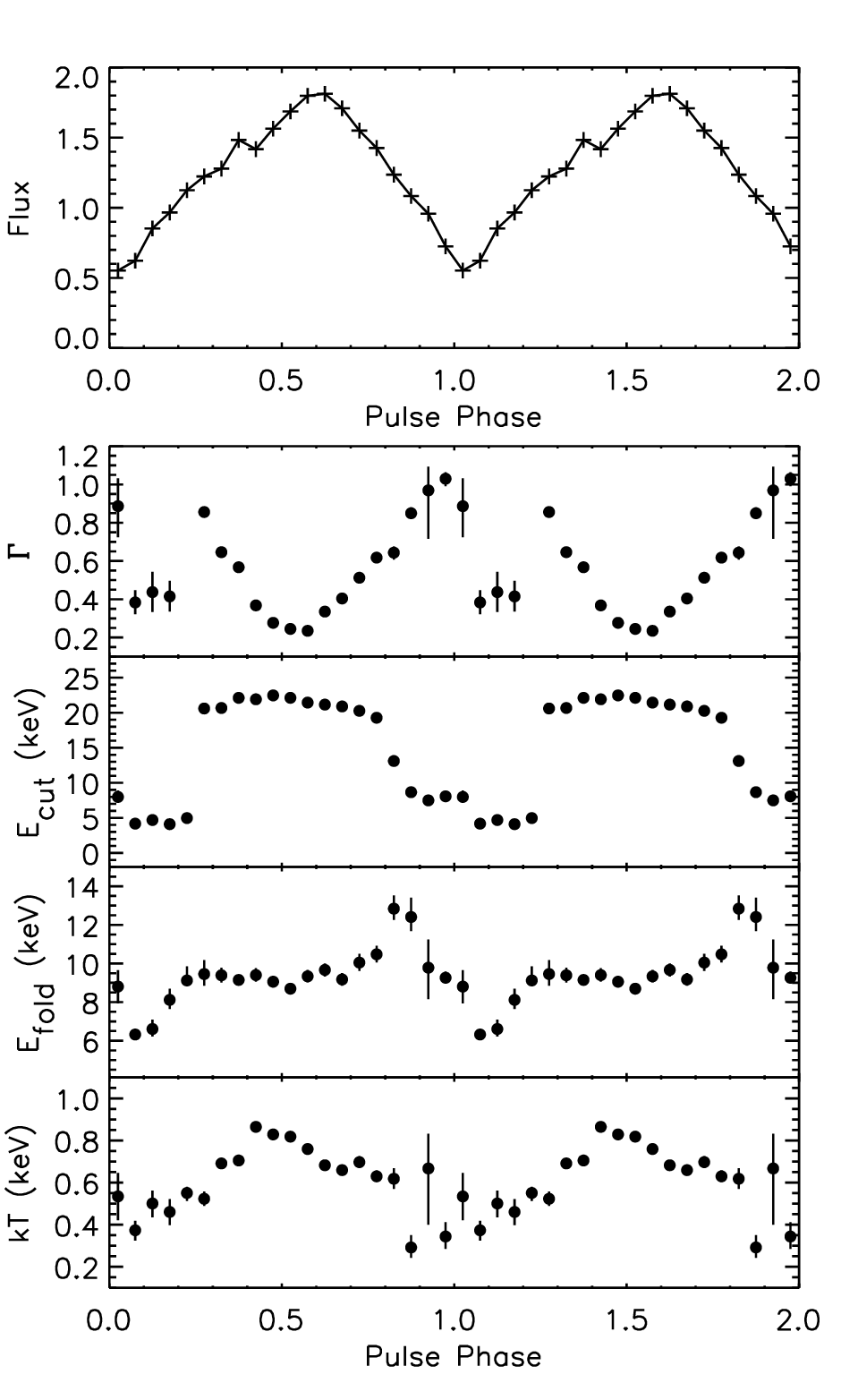} & 
\includegraphics[width=\columnwidth]{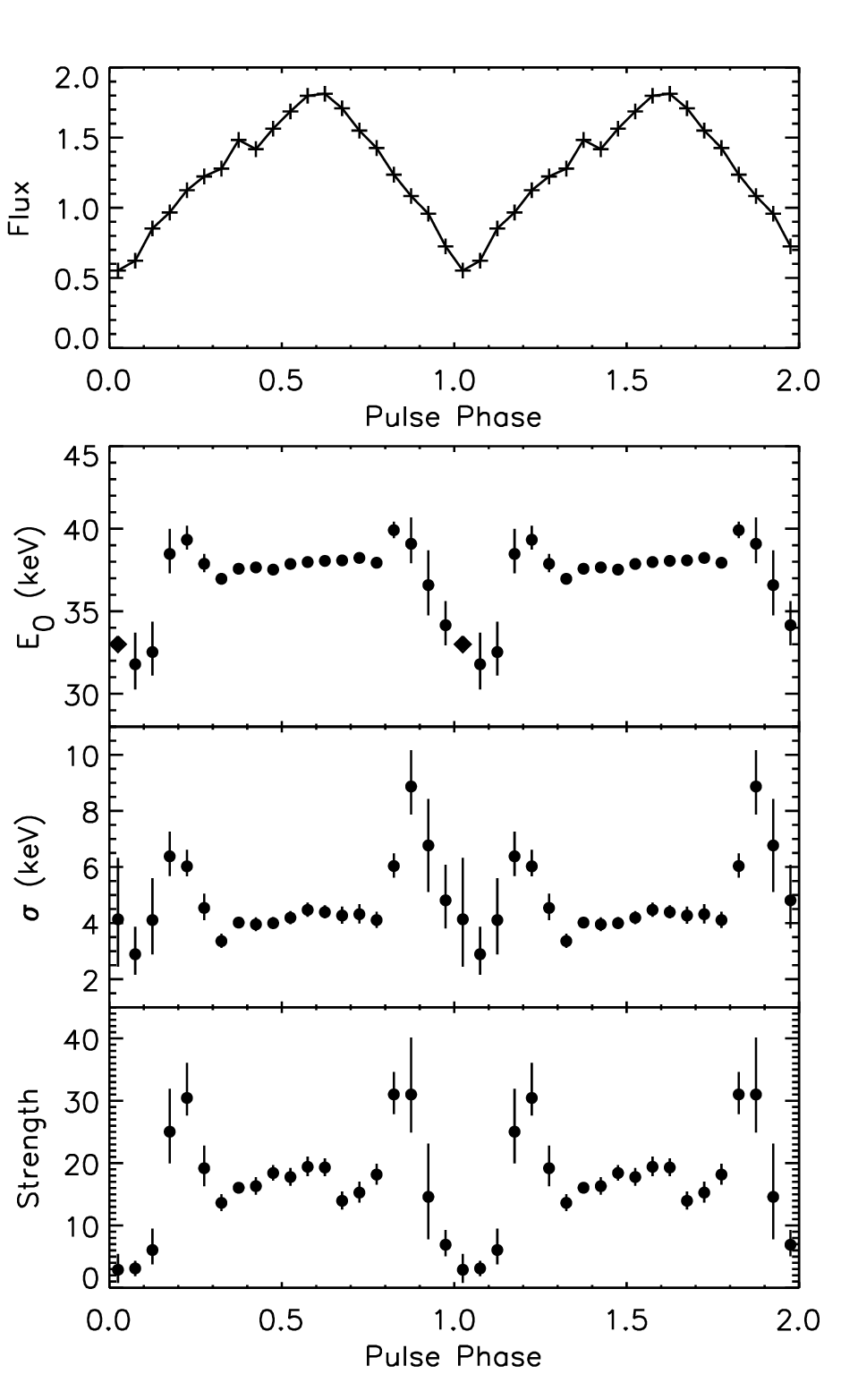} \\
\end{tabular}
\caption{Pulse-resolved spectral  parameters. Left and right  panels show
  the variations  with phase  of the continuum  and of the fundamental cyclotron  line spectral parameters, respectively. 
  The upper  panels show  the unabsorbed
  flux in  the 3--60  keV range  in units  of 10$^{-9}$  erg cm$^{-2}$
  s$^{-1}$. Error-bars at  1 $\sigma$ c.l. The diamond  symbol for the
  $E_{\rm cyc}$  parameter of the  spectrum at phase 0 indicates that
  the parameter was fixed during the fitting.}
\label{fig:pulsed}
\end{figure*}

We performed  a spin-resolved  spectral analysis of  the \emph{NuSTAR}
data, adopting a spin interval of  0.05 in phase.  This choice assures
a satisfactory  SNR for  each phase-selected  spectrum. We  are mostly
interested in  the dependence  of the cyclotron  line energy  with the
spin  phase. To  this  aim, we  simplify our  approach  and adopt  the
\textsc{highecut+bbody}     phenomenological     model,     as     the
\textsc{bwmodel}, being  an integrated version of  the total accretion
column  emission, does  not  give parameters  of  interest. The  first
harmonic is simply fitted with one \textsc{gabs} and we do not add the
possible second  harmonic, as  the line  energy is  out of  our fitted
range  (3--55 keV)  and  we are  dealing with  spectra  of much  lower
statistics than the full spin-averaged spectrum.  We verified that the
use of this model does not affect the results concerning the cyclotron
line  parameters. We  left the  equivalent absorption  column and  the
\textsc{coplref} parameters  frozen to the  spin-averaged best-fitting
values  as  we  do  not  expect any  spin-dependent  change  in  these
components.  The  best-fit values of  the continuum and  the cyclotron
line  parameters are  shown in  Fig.~\ref{fig:pulsed}.  We  found this
model to be  adequate for all the 20 spectra,  with an average reduced
$\chi^2$ of 1.02 and no significant residuals in any spectrum.

Because 50 per  cent of counts are above 8.5  keV, the flux dependence
on the  spin phase  reflects the single-peaked  folded profile  of the
high-energy  bands  as   shown  in  Fig.~\ref{fig:folded2}.  Continuum
parameters show the  highest scattering from the  average around phase
0, when the spectrum clearly shows a sudden change in the photon-index
and  in   the  cut-off  energy;  the   softer  black-body  temperature
significantly  decreases,  pointing  to   a  spin-dependence  of  this
component.

The parameters  of the  cyclotron line  also show  significant changes
when the spin phase passes through the minimum at phase 0. As the flux
decreases the  line shifts towards  lower values  as well as  the line
depth.  At the  dip bottom,  the line  is not  significantly detected,
unless  we put  some  additional constraints,  e.g.   fixing the  line
position  at the  middle between  the best-fitting  values of  the two
adjacent phase intervals.  In this case, we find a  non-zero value for
the line strength (Strength\,=\,3.5\,$\pm$\,1.5).  

\begin{figure*}
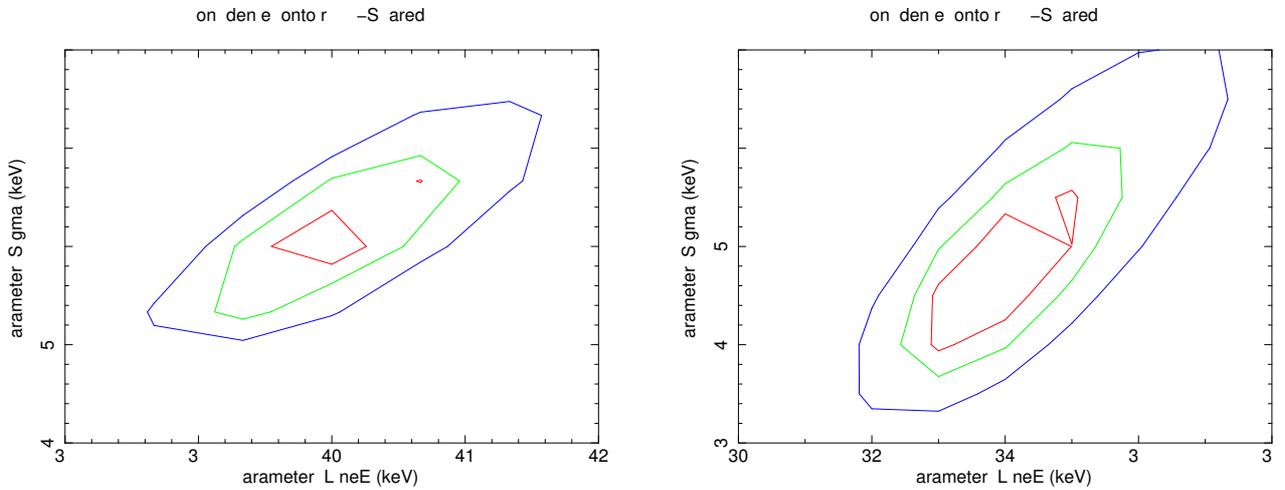

\begin{tabular}{cc}
\includegraphics[height=\columnwidth, angle=-90]{fig12a.ps} & 
\includegraphics[height=\columnwidth, angle=-90]{fig12b.ps} \\
\end{tabular}
\caption{Contour  plots for  the  cyclotron line  energy position  and
  width for  two phase-selected spectra corresponding  to interval 0.2
  (left panel)  and phase interval 0.0--0.15  (right panel). Confidence
  intervals at 68 (red contour), 90 (green contour) and 99 per cent (blue
  contour). }
\label{fig:contours}
\end{figure*}

 We  checked  if  the  intrinsic  correlation   of  the  line
  parameters affects  the measured quantities  (right  panels of
  Fig.~\ref{fig:pulsed}).    To  this   aim,  we   selected  the   most
  interesting leap  of values  that occurs  between the  first peak  of the
  folded profile at soft energies (phase $\sim$\, 0.2) and
   the  flux minimum, that independently from  the energy selection
  falls at  phase 0. Because of  the low statistics at  this phase, we
  summed  the  three  spectra  that covered  the  phase  interval  0--0.15.
 We tested the case of the line energy drop that passes from $\sim$\,40 to 
 $\sim$\,32 keV between these two phases by taking the two strongest correlated parameters, 
 the $E_{\rm line}$ and line width parameters. As shown by their  contour plots in
  Fig.~\ref{fig:contours},  the line energy error regions are well detached. 
    
\begin{figure}
\begin{center}
\includegraphics[height=\columnwidth, angle=-90]{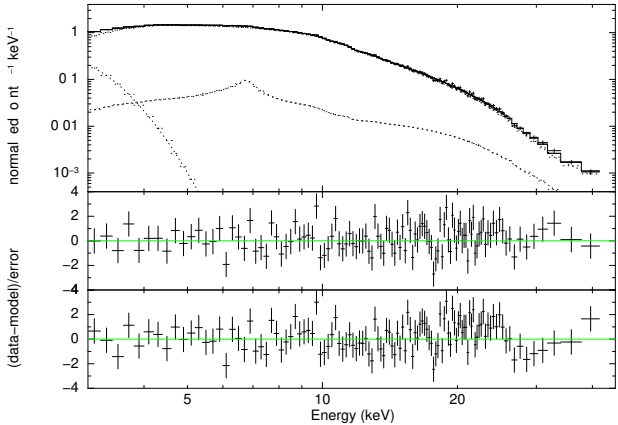}
\end{center}
\caption{\emph{NuSTAR} spectrum for phase interval 0--0.15. Residuals
  in the middle and lower panel show the continuum model with and
  without a cyclotron line at an energy of 31.6 keV. Data re-binned for clarity.}
\label{fig:dipspectrum}
\end{figure}

  In Fig.~\ref{fig:dipspectrum} we show the data and best-fit model for the
spectrum extracted from phase interval 0--0.15. The middle panel shows
residuals of  our best-fit  model continuum with  a cyclotron  line in
absorption  at  31.6\,$\pm$\,1.7  keV,  the  bottom  panel  shows  the
residuals without any cyclotron line.  There is no significant support
in the  \emph{NuSTAR} data for  any emission feature emerging  at this
particular   phase  interval   as   was  earlier   suggested  in   the
\emph{Suzaku} analysis by \citet{iwakiri12}.

Finally, it is interesting to note that the depth and the width of the
line seem to be clearly much more  affected by the shape of the softer
X-ray   spectrum,  as   these  parameters   track  more   clearly  the
double-peaked  profile.   This  is  possibly  caused  by  the  Doppler
broadening of  the line which is  caused by the thermal  motion of the
electron plasma ($kT_{\rm  e}$) at the site of the  line formation and
by  the  geometrical dependence  of  the  line  width with  the  angle
$\theta$ formed  between our  line-of-sight and  the direction  of the
electron plasma  motion according to  the broadening formula  given in
\citet{meszaros85}

\begin{equation} \label{eq:meszaros}
\left( \frac{\Delta E}{E} \right)_{FWHM} = \left( \frac{8 \textrm{ln} 2 k 
T_e}{m_e c^2}   \right)^{1/2} \textrm{cos} \theta 
\end{equation}

where    $\Delta   E$    is   the    full-width   at    half   maximum
(FWHM\,$\sim$\,2.35\,$\sigma$) of the line.

\section{Discussion}

We presented  the results  of a broadband  analysis of  X-ray spectral
data  on  the  accreting  low-mass X-ray  pulsar  \src,  covering  the
0.5--150  keV band.  For the  first time,  we apply  to its  broadband
spectrum a  self-consistent description of the  continuum emission and
of its local features. The model assumes that the channelled accretion
flow is stopped close to the NS by the radiation pressure. This is not
always the case as matter can be decelerated through other mechanisms,
e.g. Coulomb  interactions, especially at lower  accretion rates.  The
critical luminosity  above which a  radiative shock can be  formed has
been derived by \citet{basko76} and can be expressed \citep{becker07} as:

\begin{equation} \label{eq:lumcrit}
L_{\rm crit} = \frac{2.7 \times 10^{37} \sigma_{\rm T}}{\sqrt{\sigma_{\parallel} \sigma_{\bot}}} \frac{M_{\rm \star}}{M_{\odot}}
\frac{R_0}{R_{\star}} \quad \rm{erg\,s^{-1}}
\end{equation}

For the observed X-ray intensity of \src\ and assuming a distance of 9
kpc,   the  isotropic   luminosity  is   1.5\,$\times$\,10$^{37}$  erg
s$^{-1}$, which  can be in  principle below this threshold.   From the
derived  best-fitting parameters  using the  \textsc{bwmodel}, we  are
able to infer  some of these key physical parameters  in the accretion
environment.  Assuming that the  photon cross-section perpendicular to
the   B-field   is   the    same   as   the   Thomson   cross-section,
i.e. $\sigma_{\bot}$=$\sigma_{\rm T}$\,=\,6.6527\,$\times$\,10$^{-25}$
cm$^{-2}$, we can, by simple algebraic passages, derive the values for
$\alpha$,  $\sigma_{\parallel}$ and  $\bar{\sigma}$,  that are  0.335,
4.5$\times$10$^{-5}$\,$\sigma_{\rm               T}$,              and
2$\times$10$^{-4}$\,$\sigma_{\rm  T}$.   This  leads   to  an
  estimate  for  $L_{\rm crit}$  $\sim$\,1.5\,$\times$\,10$^{37}$  erg
  s$^{-1}$, that  is exactly our  estimated luminosity. We are  at the
  threshold  for the  applicability of  the radiation  dominated shock
  model, so it is reasonable to question if the derived parameters can
  be  confidently  interpreted.   To   answer  this  question,  it  is
  interesting to compare  the \textsc{bwmodel} set of  values with the
  ones obtained in a similar  study for Her X-1 \citep{wolff16}, given
  that  both  sources  accrete  through Roche  lobe  overflow  via  an
  accretion disk, and  the inferred values of the  cyclotron lines are
  similar, thus  indicating a  similar value  for the  dipole magnetic
  field. The  two sources  mainly differ in  the mass  accretion rate,
  which for  Her X-1 is higher  (2.6\,$\times$\,10$^{17}$ g s$^{-1}$),
  while  \src\  accretes  in the  range  5--14\,$\times$\,10$^{16}$  g
  s$^{-1}$.   The distance  to Her  X-1 is  well constrained  and this
  allows to  confidently set its  accretion regime above  the critical
  rate.

The  $\delta$  parameter  can  be   expressed  as  the  ratio  of  the
$y$-parameters for  the bulk  ($y_{bulk}$) and  thermal Comptonization
processes ($y_{th}$):
\begin{equation}
\delta = 4 \frac{y_{bulk}}{y_{th}} 
\end{equation}
and both sources show values  ($\delta$\,=\,2.38 in Her X-1) which are
consistent, thus  indicating that the  combined overall effect  of the
bulk and thermal Comptonizations is  similar despite the difference in
the  accretion  rate. That  the  thermal  Comptonization term  in  the
formation of  the total spectrum is  similar can also be  noted by the
electron  plasma  temperature  that  is   the  same  in  both  sources
($T_e$\,=\,4.58 keV in Her X-1).

The derived radius of the accretion column in \src\ is about a quarter
of  the  value  inferred   for  Her  X-1  ($R_0$\,=\,107\,$\pm$\,2  m)
indicating that  a much  smaller fraction of  the NS  surface actually
accretes.  The  area  of  the accreting  pole-cap  ($A_{\rm  pc}$)  is
inversely proportional to the  magnetospheric radius, $A_{\rm pc} \sim
R_{\rm NS}/2 R_{\rm mag}$, where $R_{\rm mag}$ is given by

\begin{equation} \label{magradius}
R_{\rm mag} = \kappa \times 5.1 \times 10^8 \dot{M}_{16}^{-2/7} M_{NS}^{-1/7} \mu_{30}^{4/7} \quad \rm{cm}
\end{equation}

where  $\mu_{30}$ is  the  magnetic  moment in  units  of 10$^{30}$  G
cm$^{3}$, the mass accretion rate $\dot{M}$ is in units of 10$^{16}$ g
s$^{-1}$ and the NS mass, $M_{NS}$  in units of solar masses; $\kappa$
is  the  Alfv\'en geometrical  correction  factor  that for  accretion
through a disk should be $<$\,1  and was estimated by \citet{dai15} to
be  $\sim$\,0.2. Based  on the  assumed  magnetic field  of \src\,  we
derive a  $R_{\rm mag}$  in the range  1.0--1.4 $\times$  10$^{8}$ cm,
depending on the  assumed accretion rate (or  distance).  Because $\mu
\propto B  \propto E_{\rm cyc}$, in  the case the cyclotron  lines are
produced  quite close  to the  NS  surface, it  is straightforward  to
relate  the derived  accretion  columns radii  in  these two  sources,
assuming similar  masses and radii  for the  NSs and for  the Alfv\'en
geometrical factor, as:

\begin{equation}
r_0 \propto \left( \frac{\dot{M}}{\mu^2} \right)^{1/7}.
\end{equation}

The ratio of the  column radii is then expected to be  close to 1.3, a
value apparently distant  from the factor of 4, but,  as noted before,
the  $R_0$ value  depends on  the  assumed distance,  and for  greater
distances ($>$12 kpc), we found  $R_0$ values that are compatible with
this  expected ratio.  The value  of $\sigma_{\parallel}$  is
  similar to the one obtained for Her X-1 (5.2\,$\times$\,10$^{-5}$
  $\sigma_{\rm T}$).  This parameter is  proportional to the square of
  the averaged  photon energy  ($\bar{\epsilon}$) below  the cyclotron
  energy,  and thus  implies  that the  $\bar{\epsilon}$  in \src\  is
  close to the one in Her X-1, as expected by the two similar
  spectra.  We  conclude  that the  list of  physical
  parameters derived  from the application of  the \textsc{bwmodel} to
  the broadband  spectrum of \src\  is strikingly consistent  with Her
  X-1, and thus strongly favours  a luminosity above the
  critical value of Eq.\,\ref{eq:lumcrit} and a distance
  $\geq$\,9 kpc.

\subsection{The soft X-ray black-body}

The \textsc{bwmodel} takes also into account the emission from thermal
reprocessing of the  accretion flow in the  post-shock region.  Matter
settling sub-sonically on the surface of the NS forms a \textit{mound}
of accreted  plasma and  from the derived  best-fitting values  of the
\textsc{bwmodel},  we are  able to  estimate the  temperature of  this
mound  and  the  optical  depth  at  its  top  \citep[see  Eq.\,11  in
][]{wolff16},    obtaining   6.9\,$\times$\,10$^7$    K   and    0.08,
respectively.  This temperature is greater  than the value derived for
Her  X-1 (4.3\,$\times$\,10$^7$  K), and  it is  mostly driven  by the
dependence from the accretion radius ($T\propto$ $R_0^{-2/3}$).

In our broadband  modelling, we find still evidence that  part of this
emission  could  be emitted  without  interacting  with the  accreting
plasma  in  the column  at  a  lower  temperature, with  $KT_{\rm  BB}
\sim$\,0.5 keV.  This temperature was observed to significantly change
from the spin-down to the  spin-up phase, passing from $\sim$\,0.3 keV
to  $\sim$\,0.6   keV  \citep{camero-arranz12}.   The   high  relative
fraction  of  this component  with  respect  to the  overall  observed
emission  ($\sim$\,6   per  cent),   the  small   inferred  black-body
equivalent radius, and  the spin dependence of  its parameters support
an origin from the neutron  star surface, whereas the high temperature
disfavours an origin from the truncated accretion disk, because at the
inferred inner  radius there is  not enough dissipation to  reach this
temperature.   We argue  that  this component  could  emerge from  the
surface of the NS that is not directly accreting matter, and it is not
shadowed by the accretion column.

\subsection{The shape of the cyclotron features}

Application  of the  physical  model to  the  broadband spectrum  gave
significant  residuals at  the energy  of  the cyclotron  line and  at
higher energies.  However,  when the $NuSTAR$ dataset  was analysed in
Sect.~\ref{sect:single}, the residual pattern using a phenomenological
model, appeared much flatter. It follows therefore that any constraint
on the cyclotron  shape strongly depends on the  available energy band
and on  the broadband  model used  to fit the  data. Lacking  a firmly
recognized  physical model  to fit  the cyclotron  features \citep[but
for recent    advances    in     this    direction    see
  e.g.][]{schwarm17},  we applied a  simple convolution model  of two
nested Gaussians to fit the shape of the fundamental line as also been
done by  \citet{pottschmidt05} and \citet{nakajima10} for  the case of
the accreting pulsar V0332+53. In the  case \src\ we do not detect any
asymmetry in  the residuals and we  found that these two  nested lines
had a common line energy, so that we tied them together \citep[but see
  the  case of  Cep X-4  for a  notable exception,  ][]{furst15}.  The
widths of the lines indicate a broader feature of $\sigma \sim$ 6 keV,
and a narrower one  of about half this value. With  respect to the fit
using a single Gaussian (columns 2 and 3 in Table~\ref{tab:fits}), the
line position is not changed, whereas the increase in the width of the
broader  line is  an expected  consequence  of the  flattening of  the
residual pattern on the whole fitted energy range. The position of the
line at 38 keV is consistent with previous studies \citep{orlandini98,
  coburn02,  camero-arranz12},  despite  observations  span  different
source luminosities and spin-up states.

Another  pattern of  residuals indicates  a possible  second cyclotron
harmonic but its  position is significantly different  for an expected
harmonic  ratio of  $\sim$\,2. However,  there are  both observational
evidences from  other X-ray  pulsars that  show anharmonic  ratios and
theoretical    studies   that    predict   even    larger   deviations
\citep{nishimura05,  nakajima10}.   \citet{coburn02}   did  also  note
significant  deviations  from  the  continuum at  energies  above  the
fundamental  cyclotron  using  \emph{RXTE}   data,  however  the  line
position was estimated at 80$_{-4}^{+16}$ keV, which is not consistent
with our estimate.

\subsection{The reflection component}
For the first  time we applied a  self-consistent broadband reflection
spectrum taking into account, besides  the resonant transitions of the
iron    and     low-Z    metals,    also     the    Compton-scattering
continuum. The  reflection model  was computed for  a constant
  density accretion  disk. It  represents an  angle-averaged emission,
  and  the  illuminating  source  is  assumed  central  and  isotropic
  \citep{ross93, ballantyne12}. All these  assumptions are not assured
  in the  case of  a complex  beam pattern  produced by  the accretion
  column walls of a pulsar but we retain that in this particular case,
  being the disk  truncated at more than 500  gravitational radii, our
  simplified approach might still hold.

We showed that this particular  model can well fit the high-statistics
data  of  the  \emph{NuSTAR}  data  and  the  best-fitting  parameters
indicate  a moderately  highly ionized  reflecting medium,  which well
reproduces the blended complex of \ion{Fe}{xxv} and \ion{Fe}{xxvi} and
also leaves no significant residuals at the energies of \ion{O}{viii},
but an excess around the Ne complex is still evident, and possibly due
to    an   over-abundance    of    this    element.    The    combined
\emph{NuSTAR}/\emph{Swift} observation is the  first to constrain iron
resonant   features   in  this   high   ionisation   state,  as   past
\emph{Chandra} and \emph{Suzaku} observations found iron in a neutral,
or mildly ionized state. This can  be due to the higher accretion rate
of the source during the  \emph{NuSTAR} observation, which at the same
time, brings the accretion disk closer to the compact object and gives
also a stronger irradiating flux.  We note that this is also supported
by the  measured higher equivalent  widths, as  the iron line  seen by
\emph{Chandra} in  2010 had a width  of 36\,$\pm$\,15 eV and  an EW of
18\,$\pm$\,6 eV \citep{koliopanos16}.

We  calculated  a  reflection  strength of  $\sim$\,0.05  (defined  as
the ratio in the 20--40  keV range of the reflection component
  flux to the continuum). The  reflection fraction, estimated with the
  \textsc{pexriv} model fitting only the reflection bump, results in a
  similar  value  \citep[see][for  the   difference  between  the  two
    definitions]{dauser16}.   This value  is considerably  lower with
respect  to the  non-pulsating NS  X-ray binaries,  since the  disk is
truncated at a larger distance from the NS and, thus, subtends a lower
angle from the  primary source of hard X-ray photons.   In the case of
GRO  J1744-28, a  similar low-mass  X-ray pulsar  seen at  similar low
inclination  angle,   the  inferred   reflection  fraction   was  also
$\sim$\,0.05,  but  the  width  of  the  reflecting  lines  was  found
significantly  wider.  This  indicates  that the  reflecting disk  was
similar in  size but the inner  radius was much closer  to the pulsar,
where Keplerian and relativistic effects to the line broadening become
statistically detectable in the spectrum \citep{younes15, dai15}.

\section{Acknowledgments}
AD  acknowledges contract  ASI-INAF  I/004/11/0.   We acknowledge  the
\textsc{heasoft}  software use  developed and  maintained by  the NASA
High   Energy    Astrophysics   Science   Archive    Research   Center
[\textsc{heasarc}]).

\bibliographystyle{mnras}
\bibliography{refs}

\begin{thebibliography}{}
\makeatletter
\relax
\def\mn@urlcharsother{\let\do\@makeother \do\$\do\&\do\#\do\^\do\_\do\%\do\~}
\def\mn@doi{\begingroup\mn@urlcharsother \@ifnextchar [ {\mn@doi@}
  {\mn@doi@[]}}
\def\mn@doi@[#1]#2{\def\@tempa{#1}\ifx\@tempa\@empty \href
  {http://dx.doi.org/#2} {doi:#2}\else \href {http://dx.doi.org/#2} {#1}\fi
  \endgroup}
\def\mn@eprint#1#2{\mn@eprint@#1:#2::\@nil}
\def\mn@eprint@arXiv#1{\href {http://arxiv.org/abs/#1} {{\tt arXiv:#1}}}
\def\mn@eprint@dblp#1{\href {http://dblp.uni-trier.de/rec/bibtex/#1.xml}
  {dblp:#1}}
\def\mn@eprint@#1:#2:#3:#4\@nil{\def\@tempa {#1}\def\@tempb {#2}\def\@tempc
  {#3}\ifx \@tempc \@empty \let \@tempc \@tempb \let \@tempb \@tempa \fi \ifx
  \@tempb \@empty \def\@tempb {arXiv}\fi \@ifundefined
  {mn@eprint@\@tempb}{\@tempb:\@tempc}{\expandafter \expandafter \csname
  mn@eprint@\@tempb\endcsname \expandafter{\@tempc}}}

\bibitem[\protect\citeauthoryear{{An} et~al.,}{{An} et~al.}{2015}]{an15}
{An} H.,  et~al., 2015, \mn@doi [\apj] {10.1088/0004-637X/807/1/93}, \href
  {http://adsabs.harvard.edu/abs/2015ApJ...807...93A} {807, 93}

\bibitem[\protect\citeauthoryear{{Ballantyne}, {Purvis}, {Strausbaugh}  \&
  {Hickox}}{{Ballantyne} et~al.}{2012}]{ballantyne12}
{Ballantyne} D.~R.,  {Purvis} J.~D.,  {Strausbaugh} R.~G.,   {Hickox} R.~C.,
  2012, \mn@doi [\apjl] {10.1088/2041-8205/747/2/L35}, \href
  {http://adsabs.harvard.edu/abs/2012ApJ...747L..35B} {747, L35}

\bibitem[\protect\citeauthoryear{{Barthelmy}, {Barbier}  \&
  {Cummings}}{{Barthelmy} et~al.}{2005}]{barthelmy05}
{Barthelmy} S.~D.,  {Barbier} L.~M.,   {Cummings} J.~R. e.~a.,  2005, \mn@doi
  [\ssr] {10.1007/s11214-005-5096-3}, \href
  {http://adsabs.harvard.edu/abs/2005SSRv..120..143B} {120, 143}

\bibitem[\protect\citeauthoryear{{Basko} \& {Sunyaev}}{{Basko} \&
  {Sunyaev}}{1976}]{basko76}
{Basko} M.~M.,  {Sunyaev} R.~A.,  1976, \mnras, \href
  {http://adsabs.harvard.edu/abs/1976MNRAS.175..395B} {175, 395}

\bibitem[\protect\citeauthoryear{{Baumgartner}, {Tueller}, {Markwardt},
  {Skinner}, {Barthelmy}, {Mushotzky}, {Evans}  \& {Gehrels}}{{Baumgartner}
  et~al.}{2013}]{baumgartner13}
{Baumgartner} W.~H.,  {Tueller} J.,  {Markwardt} C.~B.,  {Skinner} G.~K.,
  {Barthelmy} S.,  {Mushotzky} R.~F.,  {Evans} P.~A.,   {Gehrels} N.,  2013,
  \mn@doi [\apjs] {10.1088/0067-0049/207/2/19}, \href
  {http://adsabs.harvard.edu/abs/2013ApJS..207...19B} {207, 19}

\bibitem[\protect\citeauthoryear{{Becker} \& {Wolff}}{{Becker} \&
  {Wolff}}{2007}]{becker07}
{Becker} P.~A.,  {Wolff} M.~T.,  2007, \mn@doi [\apj] {10.1086/509108}, \href
  {http://adsabs.harvard.edu/abs/2007ApJ...654..435B} {654, 435}

\bibitem[\protect\citeauthoryear{{Beri}, {Jain}, {Paul}  \& {Raichur}}{{Beri}
  et~al.}{2014}]{beri14}
{Beri} A.,  {Jain} C.,  {Paul} B.,   {Raichur} H.,  2014, \mn@doi [\mnras]
  {10.1093/mnras/stu087}, \href
  {http://adsabs.harvard.edu/abs/2014MNRAS.439.1940B} {439, 1940}

\bibitem[\protect\citeauthoryear{{Burrows}, {Hill}, {Nousek}, {Kennea}, {Wells}
   \& {Osborne}}{{Burrows} et~al.}{2005}]{burrows05}
{Burrows} D.~N.,  {Hill} J.~E.,  {Nousek} J.~A.,  {Kennea} J.~A.,  {Wells} A.,
   {Osborne} 2005, \mn@doi [Space Science Reviews] {10.1007/s11214-005-5097-2},
  \href {http://adsabs.harvard.edu/abs/2005SSRv..120..165B} {120, 165}

\bibitem[\protect\citeauthoryear{{Camero-Arranz}, {Pottschmidt}, {Finger},
  {Ikhsanov}, {Wilson-Hodge}  \& {Marcu}}{{Camero-Arranz}
  et~al.}{2012}]{camero-arranz12}
{Camero-Arranz} A.,  {Pottschmidt} K.,  {Finger} M.~H.,  {Ikhsanov} N.~R.,
  {Wilson-Hodge} C.~A.,   {Marcu} D.~M.,  2012, \mn@doi [\aap]
  {10.1051/0004-6361/201219656}, \href
  {http://adsabs.harvard.edu/abs/2012A%26A...546A..40C} {546, A40}

\bibitem[\protect\citeauthoryear{{Chakrabarty}}{{Chakrabarty}}{1998}]{chakrabarty98}
{Chakrabarty} D.,  1998, \mn@doi [\apj] {10.1086/305035}, \href
  {http://adsabs.harvard.edu/abs/1998ApJ...492..342C} {492, 342}

\bibitem[\protect\citeauthoryear{{Coburn}, {Heindl}, {Rothschild}, {Gruber},
  {Kreykenbohm}, {Wilms}, {Kretschmar}  \& {Staubert}}{{Coburn}
  et~al.}{2002}]{coburn02}
{Coburn} W.,  {Heindl} W.~A.,  {Rothschild} R.~E.,  {Gruber} D.~E.,
  {Kreykenbohm} I.,  {Wilms} J.,  {Kretschmar} P.,   {Staubert} R.,  2002,
  \mn@doi [\apj] {10.1086/343033}, \href
  {http://adsabs.harvard.edu/abs/2002ApJ...580..394C} {580, 394}

\bibitem[\protect\citeauthoryear{{D'A{\`i}} et~al.,}{{D'A{\`i}}
  et~al.}{2015}]{dai15}
{D'A{\`i}} A.,  et~al., 2015, \mn@doi [\mnras] {10.1093/mnras/stv531}, \href
  {http://adsabs.harvard.edu/abs/2015MNRAS.449.4288D} {449, 4288}

\bibitem[\protect\citeauthoryear{{Dauser}, {Garc{\'{\i}}a}, {Walton},
  {Eikmann}, {Kallman}, {McClintock}  \& {Wilms}}{{Dauser}
  et~al.}{2016}]{dauser16}
{Dauser} T.,  {Garc{\'{\i}}a} J.,  {Walton} D.~J.,  {Eikmann} W.,  {Kallman}
  T.,  {McClintock} J.,   {Wilms} J.,  2016, \mn@doi [\aap]
  {10.1051/0004-6361/201628135}, \href
  {http://adsabs.harvard.edu/abs/2016A%26A...590A..76D} {590, A76}

\bibitem[\protect\citeauthoryear{{Doroshenko}, {Tsygankov}, {Mushtukov},
  {Lutovinov}, {Santangelo}, {Suleimanov}  \& {Poutanen}}{{Doroshenko}
  et~al.}{2017}]{doroshenko17}
{Doroshenko} V.,  {Tsygankov} S.~S.,  {Mushtukov} A.~A.,  {Lutovinov} A.~A.,
  {Santangelo} A.,  {Suleimanov} V.~F.,   {Poutanen} J.,  2017, \mn@doi
  [\mnras] {10.1093/mnras/stw3236}, \href
  {http://adsabs.harvard.edu/abs/2017MNRAS.466.2143D} {466, 2143}

\bibitem[\protect\citeauthoryear{{Ferrigno}, {Becker}, {Segreto}, {Mineo}  \&
  {Santangelo}}{{Ferrigno} et~al.}{2009}]{ferrigno09}
{Ferrigno} C.,  {Becker} P.~A.,  {Segreto} A.,  {Mineo} T.,   {Santangelo} A.,
  2009, \mn@doi [\aap] {10.1051/0004-6361/200809373}, \href
  {http://adsabs.harvard.edu/abs/2009A%26A...498..825F} {498, 825}

\bibitem[\protect\citeauthoryear{{Ferrigno} et~al.,}{{Ferrigno}
  et~al.}{2016}]{ferrigno16}
{Ferrigno} C.,  et~al., 2016, \mn@doi [\aap] {10.1051/0004-6361/201628865},
  \href {http://adsabs.harvard.edu/abs/2016A%26A...595A..17F} {595, A17}

\bibitem[\protect\citeauthoryear{{F{\"u}rst} et~al.,}{{F{\"u}rst}
  et~al.}{2013}]{furst13}
{F{\"u}rst} F.,  et~al., 2013, \mn@doi [\apj] {10.1088/0004-637X/779/1/69},
  \href {http://adsabs.harvard.edu/abs/2013ApJ...779...69F} {779, 69}

\bibitem[\protect\citeauthoryear{{F{\"u}rst} et~al.,}{{F{\"u}rst}
  et~al.}{2015}]{furst15}
{F{\"u}rst} F.,  et~al., 2015, \mn@doi [\apjl] {10.1088/2041-8205/806/2/L24},
  \href {http://adsabs.harvard.edu/abs/2015ApJ...806L..24F} {806, L24}

\bibitem[\protect\citeauthoryear{{Harrison} et~al.,}{{Harrison}
  et~al.}{2013}]{harrison13}
{Harrison} F.~A.,  et~al., 2013, \mn@doi [\apj] {10.1088/0004-637X/770/2/103},
  \href {http://adsabs.harvard.edu/abs/2013ApJ...770..103H} {770, 103}

\bibitem[\protect\citeauthoryear{{Iwakiri} et~al.,}{{Iwakiri}
  et~al.}{2012}]{iwakiri12}
{Iwakiri} W.~B.,  et~al., 2012, \mn@doi [\apj] {10.1088/0004-637X/751/1/35},
  \href {http://adsabs.harvard.edu/abs/2012ApJ...751...35I} {751, 35}

\bibitem[\protect\citeauthoryear{{Jain}, {Paul}  \& {Dutta}}{{Jain}
  et~al.}{2010}]{jain10}
{Jain} C.,  {Paul} B.,   {Dutta} A.,  2010, \mn@doi [\mnras]
  {10.1111/j.1365-2966.2009.16170.x}, \href
  {http://adsabs.harvard.edu/abs/2010MNRAS.403..920J} {403, 920}

\bibitem[\protect\citeauthoryear{{Kii}, {Hayakawa}, {Nagase}, {Ikegami}  \&
  {Kawai}}{{Kii} et~al.}{1986}]{kii86}
{Kii} T.,  {Hayakawa} S.,  {Nagase} F.,  {Ikegami} T.,   {Kawai} N.,  1986,
  \pasj, \href {http://adsabs.harvard.edu/abs/1986PASJ...38..751K} {38, 751}

\bibitem[\protect\citeauthoryear{{Koliopanos} \& {Gilfanov}}{{Koliopanos} \&
  {Gilfanov}}{2016}]{koliopanos16}
{Koliopanos} F.,  {Gilfanov} M.,  2016, \mn@doi [\mnras]
  {10.1093/mnras/stv2873}, \href
  {http://adsabs.harvard.edu/abs/2016MNRAS.456.3535K} {456, 3535}

\bibitem[\protect\citeauthoryear{{Krauss}, {Schulz}, {Chakrabarty}, {Juett}  \&
  {Cottam}}{{Krauss} et~al.}{2007}]{krauss07}
{Krauss} M.~I.,  {Schulz} N.~S.,  {Chakrabarty} D.,  {Juett} A.~M.,   {Cottam}
  J.,  2007, \mn@doi [\apj] {10.1086/513592}, \href
  {http://adsabs.harvard.edu/abs/2007ApJ...660..605K} {660, 605}

\bibitem[\protect\citeauthoryear{{La Parola}, {Cusumano}, {Segreto}  \&
  {D'A{\`i}}}{{La Parola} et~al.}{2016}]{laparola16}
{La Parola} V.,  {Cusumano} G.,  {Segreto} A.,   {D'A{\`i}} A.,  2016, \mn@doi
  [\mnras] {10.1093/mnras/stw1915}, \href
  {http://adsabs.harvard.edu/abs/2016MNRAS.463..185L} {463, 185}

\bibitem[\protect\citeauthoryear{{Meszaros} \& {Nagel}}{{Meszaros} \&
  {Nagel}}{1985}]{meszaros85}
{Meszaros} P.,  {Nagel} W.,  1985, \mn@doi [\apj] {10.1086/163594}, \href
  {http://adsabs.harvard.edu/abs/1985ApJ...298..147M} {298, 147}

\bibitem[\protect\citeauthoryear{{Middleditch}, {Mason}, {Nelson}  \&
  {White}}{{Middleditch} et~al.}{1981}]{middleditch81}
{Middleditch} J.,  {Mason} K.~O.,  {Nelson} J.~E.,   {White} N.~E.,  1981,
  \mn@doi [\apj] {10.1086/158772}, \href
  {http://adsabs.harvard.edu/abs/1981ApJ...244.1001M} {244, 1001}

\bibitem[\protect\citeauthoryear{{Nakajima}, {Mihara}  \&
  {Makishima}}{{Nakajima} et~al.}{2010}]{nakajima10}
{Nakajima} M.,  {Mihara} T.,   {Makishima} K.,  2010, \mn@doi [\apj]
  {10.1088/0004-637X/710/2/1755}, \href
  {http://adsabs.harvard.edu/abs/2010ApJ...710.1755N} {710, 1755}

\bibitem[\protect\citeauthoryear{{Nishimura}}{{Nishimura}}{2005}]{nishimura05}
{Nishimura} O.,  2005, \mn@doi [\pasj] {10.1093/pasj/57.5.769}, \href
  {http://adsabs.harvard.edu/abs/2005PASJ...57..769N} {57, 769}

\bibitem[\protect\citeauthoryear{{Orlandini} et~al.,}{{Orlandini}
  et~al.}{1998}]{orlandini98}
{Orlandini} M.,  et~al., 1998, \mn@doi [\apjl] {10.1086/311404}, \href
  {http://adsabs.harvard.edu/abs/1998ApJ...500L.163O} {500, L163}

\bibitem[\protect\citeauthoryear{{Pottschmidt} et~al.,}{{Pottschmidt}
  et~al.}{2005}]{pottschmidt05}
{Pottschmidt} K.,  et~al., 2005, \mn@doi [\apjl] {10.1086/498689}, \href
  {http://adsabs.harvard.edu/abs/2005ApJ...634L..97P} {634, L97}

\bibitem[\protect\citeauthoryear{{Rappaport}, {Markert}, {Li}, {Clark},
  {Jernigan}  \& {McClintock}}{{Rappaport} et~al.}{1977}]{rappaport77}
{Rappaport} S.,  {Markert} T.,  {Li} F.~K.,  {Clark} G.~W.,  {Jernigan} J.~G.,
   {McClintock} J.~E.,  1977, \mn@doi [\apjl] {10.1086/182532}, \href
  {http://ads.ari.uni-heidelberg.de/abs/1977ApJ...217L..29R} {217, L29}

\bibitem[\protect\citeauthoryear{{Ross} \& {Fabian}}{{Ross} \&
  {Fabian}}{1993}]{ross93}
{Ross} R.~R.,  {Fabian} A.~C.,  1993, \mn@doi [\mnras]
  {10.1093/mnras/261.1.74}, \href
  {http://adsabs.harvard.edu/abs/1993MNRAS.261...74R} {261, 74}

\bibitem[\protect\citeauthoryear{{Schulz}, {Chakrabarty}, {Marshall},
  {Canizares}, {Lee}  \& {Houck}}{{Schulz} et~al.}{2001}]{schulz01}
{Schulz} N.~S.,  {Chakrabarty} D.,  {Marshall} H.~L.,  {Canizares} C.~R.,
  {Lee} J.~C.,   {Houck} J.,  2001, \mn@doi [\apj] {10.1086/323988}, \href
  {http://adsabs.harvard.edu/abs/2001ApJ...563..941S} {563, 941}

\bibitem[\protect\citeauthoryear{{Schwarm} et~al.,}{{Schwarm}
  et~al.}{2017}]{schwarm17}
{Schwarm} F.-W.,  et~al., 2017, \mn@doi [\aap] {10.1051/0004-6361/201629352},
  \href {http://adsabs.harvard.edu/abs/2017A%26A...597A...3S} {597, A3}

\bibitem[\protect\citeauthoryear{{Segreto}, {Cusumano}, {Ferrigno}, {La
  Parola}, {Mangano}, {Mineo}  \& {Romano}}{{Segreto} et~al.}{2010}]{segreto10}
{Segreto} A.,  {Cusumano} G.,  {Ferrigno} C.,  {La Parola} V.,  {Mangano} V.,
  {Mineo} T.,   {Romano} P.,  2010, \mn@doi [\aap]
  {10.1051/0004-6361/200911779}, \href
  {http://adsabs.harvard.edu/abs/2010A%26A...510A..47S} {510, A47}

\bibitem[\protect\citeauthoryear{{Takagi}, {Mihara}, {Sugizaki}, {Makishima}
  \& {Morii}}{{Takagi} et~al.}{2016}]{takagi16}
{Takagi} T.,  {Mihara} T.,  {Sugizaki} M.,  {Makishima} K.,   {Morii} M.,
  2016, \mn@doi [\pasj] {10.1093/pasj/psw010}, \href
  {http://adsabs.harvard.edu/abs/2016PASJ...68S..13T} {68, S13}

\bibitem[\protect\citeauthoryear{{Wolff} et~al.,}{{Wolff}
  et~al.}{2016}]{wolff16}
{Wolff} M.~T.,  et~al., 2016, \mn@doi [\apj] {10.3847/0004-637X/831/2/194},
  \href {http://adsabs.harvard.edu/abs/2016ApJ...831..194W} {831, 194}

\bibitem[\protect\citeauthoryear{{Younes}, {Kouveliotou}  \&
  {Grefenstette}}{{Younes} et~al.}{2015}]{younes15}
{Younes} G.,  {Kouveliotou} C.,   {Grefenstette} B.~W.,  2015, \mn@doi [\apj]
  {10.1088/0004-637X/804/1/43}, \href
  {http://adsabs.harvard.edu/abs/2015ApJ...804...43Y} {804, 43}

\makeatother
\end{thebibliography}
\bsp
\label{lastpage}
\end{document}